\documentclass[lettersize,journal]{IEEEtran}
\usepackage{amsmath,amsfonts}
\usepackage{algorithmic}
\usepackage{array}
\usepackage{textcomp}
\usepackage{stfloats}
\usepackage{url}
\usepackage{verbatim}
\usepackage{graphicx}
\usepackage{caption}
\usepackage{subcaption}
\usepackage{balance}
\usepackage{booktabs}
\usepackage{dsfont}
\usepackage{bm}
\usepackage{color}
\usepackage{setspace}
\usepackage{multirow}
\usepackage{cite}
\def\BibTeX{{\rm B\kern-.05em{\sc i\kern-.025em b}\kern-.08em
    T\kern-.1667em\lower.7ex\hbox{E}\kern-.125emX}}

\begin{document}
\title{Variational Bayesian Adaptive Learning of Deep Latent Variables for Acoustic Knowledge Transfer}

\author{Hu Hu$^{1}$,
      Sabato Marco Siniscalchi$^{1,2,3}$~\IEEEmembership{Senior Member,~IEEE},
      Chao-Han Huck Yang$^{4}$~\IEEEmembership{Student Member,~IEEE} and
      Chin-Hui Lee$^{1}$~\IEEEmembership{Fellow,~IEEE}
\thanks{Hu Hu and Chin-Hui Lee are with the School of Electrical and Computer Engineering, Georgia Institute of Technology, Atlanta, GA, USA (e-mail: hhu96@gatech.edu, chl@ece.gatech.edu), S. M. Siniscalchi is with the University of Palermo, Georgia Institute of Technology, and Norwegian University of Science and Technology (e-mail: sabatomarco.siniscalchi@unipa.it), and Chao-Han Huck Yang is with Nvidia (e-mail: hucky@nvidia.com)}}

{}

\maketitle

\begin{abstract}

In this work, we propose a novel variational Bayesian adaptive learning approach for cross-domain knowledge transfer to address acoustic mismatches between training and testing conditions, such as recording devices and environmental noise. Different from the traditional Bayesian approaches that impose uncertainties on model parameters risking the curse of dimensionality due to the huge number of parameters, we focus on estimating a manageable number of latent variables in deep neural models. Knowledge learned from a source domain is thus encoded in prior distributions of deep latent variables and optimally combined, in a Bayesian sense, with a small set of adaptation data from a target domain to approximate the corresponding posterior distributions. Two different strategies are proposed and investigated to estimate the posterior distributions: Gaussian mean-field variational inference, and empirical Bayes. These strategies address the presence or absence of parallel data in the source and target domains. Furthermore, structural relationship modeling is investigated to enhance the approximation. We evaluated our proposed approaches on two acoustic adaptation tasks: 1) device adaptation for acoustic scene classification, and 2) noise adaptation for spoken command recognition. Experimental results show that the proposed variational Bayesian adaptive learning approach can obtain good improvements on target domain data, and consistently outperforms state-of-the-art knowledge transfer methods.

\end{abstract}


\section{Introduction}

Recent advancements in model-based classification applications, particularly for acoustic models, owe much to the synergistic evolution of deep neural networks (DNNs) and the proliferation of extensive datasets. These developments have propelled state-of-the-art achievements in acoustic system design, leveraging the computational power of DNNs and the rich variability captured in large-scale data collections \cite{hinton2012deep, seide2011conversational, xu2014regression, chan2016listen, radford2023robust}.
The efficacy of these data-driven solutions, however, is intricately linked to the representativeness of the training data concerning the statistical variations encountered in real-world testing environments. Nonetheless, models concerning acoustic information are particularly susceptible to performance degradation due to mismatches related to variations in speakers, recording devices, and ambient noise conditions.


Acoustic scene classification (ASC) exemplifies such challenges, as device mismatch in production is a persistent issue \cite{mesaros2018multi, gharib2018unsupervised, koutini2020low, sonowal2022novel}. To address the latter issue, researchers usually pool diverse datasets to forge more robust systems \cite{hu2020device, kim2022domain, sonowal2022novel, morocutti2023device, jiang2023multi}. Nonetheless, gathering and annotating domain-specific data remains a costly and time-consuming endeavor. As long as the mismatch between source and target domains is inevitable, most data-driven models need to be rebuilt from scratch or carefully fine-tuned by using newly collected data.
Thus, addressing these mismatches necessitates robust adaptation strategies to transfer knowledge from a well-characterized source domain to less familiar target domains, especially when adaptation data is limited. The development of efficient adaptation techniques is thus critical. Those techniques must leverage limited target domain data to refine existing models, mitigating the risks of catastrophic forgetting \cite{goodfellow2013empirical, huang2015maximum, kirkpatrick2017overcoming} and the curse of dimensionality \cite{powell2007approximate, poggio2017and}. 

Bayesian adaptive learning is an optimal way to combine prior knowledge and a set of new adaptation data. Therefore, it is often adopted to deal with inconsistencies between training and testing conditions and to transfer knowledge across domains. This approach harnesses a mathematical framework to model uncertainties and integrate prior knowledge effectively, thereby enhancing model adaptation capabilities \cite{ripley1996pattern, lee2000adaptive}.
Through Bayesian formulations, it becomes possible to adjust model parameters optimally based on posterior distributions, offering a refined mechanism for addressing acoustic mismatches.
Traditional Bayesian formulations usually impose uncertainties on model parameters and focus on obtaining point estimates \cite{lee2000adaptive, gauvain1994maximum} of the unknown parameters of the posterior distributions \cite{huang2015maximum, huang2017bayesian}.
However, the number of DNN parameters is usually much larger than the available adaptation samples, causing the estimated DNN parameters to be inaccurate and making Bayesian adaptation less effective.

In this work, we propose a novel variational Bayesian (VB) knowledge transfer (VBKT) framework to model the distributions of deep latent variables for effective cross-domain adaptation. Instead of carrying out traditional Bayesian estimation on model parameters, we focus on estimating a manageable number of DNN latent variables.
The deep latent variables here refer to the unobservable representations of data, and they usually correspond to intermediate hidden embedding outputs from a specific layer of DNNs.
In particular, by leveraging upon variational inference, the distributions of the source latent variables (prior) are combined with the knowledge learned from target data (likelihood) to yield the distributions of the target latent variables (posterior). Prior knowledge from the source domain is thus encoded and transferred to the target domain, by approximating the posterior distributions of latent variables.
To accomplish model transfer, knowledge learned from a source domain is encoded in prior distributions of latent variables and optimally combined, in a Bayesian sense, with a small set of adaptation data from a target domain to approximate the corresponding posterior distributions.
We present two different prior approximation mechanisms, namely Gaussian mean-field variational inference (GMFVI) and Empirical Bayes (EB), depending on the availability of parallel data across source and target domains.
We also explore structural relationship modeling to enhance the approximation.
Experimental results on: (i) device adaptation in acoustic scene classification, and (ii) noise adaptation in spoken command recognition, show the superiority of our variational Bayesian adaptive learning framework over existing adaptation techniques available in the literature.

The rest of this paper is organized as follows: Section \ref{sec:related} discusses background and related work. Section~\ref{sec:bayes} introduces the proposed Bayesian adaptive learning framework, which focuses on estimating deep latent variables.  Section~\ref{sec:vbkt} describes the proposed variational Bayesian adaptive learning approach, including the derivation and two estimation solutions. Section~\ref{sec:exp} presents our experimental results and reports our analysis. Finally, Section~\ref{sec:con} concludes our work.

\section{Related Works}
\label{sec:related}

Adaptation techniques are effective solutions to combat the mismatch problem by utilizing a limited amount of target data to update the parameters of an existing source model to new conditions. Among adaptation techniques, Bayesian adaptative learning provides a mathematical framework to model uncertainties and incorporate prior knowledge for cross-domain knowledge transfer.
The Bayesian adaptive learning framework is known to facilitate constructing an adaptive system for specific target conditions in a particular environment. Thus the mismatch between training and testing can be reduced with an overall improvement of system performance.

A common Bayesian adaptation mechanism is the maximum a posteriori (MAP) estimation \cite{duda1973pattern, ripley1996pattern} of parameters in the posterior distributions, which amounts to maximizing the posterior probability and gives us a point estimation of the model parameters. It has been proven effective in handling acoustic mismatches in hidden Markov models (HMMs) \cite{lee2000adaptive, gauvain1994maximum, siohan2001joint} and deep models \cite{huang2015maximum, huang2017bayesian, kirkpatrick2017overcoming} by assuming a distribution on the model parameters.
Online Bayesian learning \cite{huo1997line} and joint Bayesian learning of HMM structures and transformation parameters have also been extensively studied \cite{shinoda2001structural, siohan2001joint, huang2017hierarchical}.
Besides point estimation, the entire posterior distributions can also be approximated by other Bayesian approaches, such as Markov Chain Monte Carlo (MCMC) \cite{neal1993probabilistic}, assumed density filtering \cite{maybeck1982stochastic}, and stochastic variational Bayes (VB) \cite{beal2003variational, paisley2012variational}.
The VB approach, in particular, performs an estimation on the entire posterior distribution via a stochastic variational inference method and transforms an estimation problem into an optimization one that can be solved numerically by leveraging a variational distribution. It has been applied to both conventional HMM-based acoustic models \cite{watanabe2004variational} and deep models \cite{kingma2013auto, hu2022variational, si2021variational}.

Traditional Bayesian adaptive learning methods usually impose uncertainties on model parameters. That works well to model uncertainties on classical machine learning models with a limited amount of parameters. However, accurately estimating distributions for DNN parameters is much more complicated due to the large amount of model parameters.
Thus, the most recent popular knowledge transfer frameworks for DNN models are usually feature-based - that is, a latent variable-based point estimation from a Bayesian perspective, such as the teacher-student learning (TSL) technique (also known as knowledge distillation) \cite{li2014learning, hinton2015distilling}. TSL transfers knowledge learned by the source/teacher model and encoded in its softened outputs (model outputs after softmax), to the target/student model, where the target model directly mimics the final prediction of the source model through a KL divergence loss.
The idea was later extended to the hidden embedding of intermediate layers \cite{romero2014fitnets, zagoruyko2016paying, heo2019comprehensive}, where different embedded representations are proposed to encode and transfer knowledge. However, instead of considering the whole distribution of latent variables, those methods typically rely on point estimations, potentially leading to sub-optimal results and loss of distributional information.

In our previous work \cite{hu2022variational}, a variational Bayesian (VB) approach was proposed to learn distributions of latent variables in DNN models for cross-domain knowledge transfer. In this work, we build upon our previous effort by extending and enhancing our methodology. Specifically, we introduce two strategies to perform the transfer learning, namely Gaussian mean-field variational inference,
and empirical Bayes. Furthermore, we also explore the possibility of modeling the structural relationship and thereby be able to (i) deal with different data conditions, and (ii) enhance the approximation. The details of the proposed variational Bayesian adaptive learning framework are given in the following sections.

\begin{figure*}[t]
  \centering
  \vspace{0.2cm}
  \includegraphics[width=0.97\linewidth]{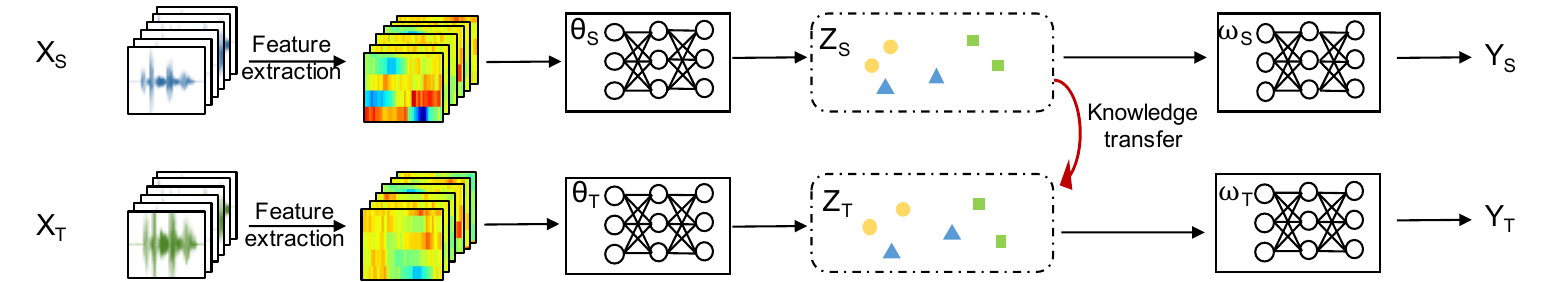}
  \vspace{0.3cm}
  \caption{Illustration of the deep latent variables based acoustic knowledge transfer framework.}
  \label{fig:framework}
\end{figure*}

\section{Bayesian Adaptation on Deep Latent Variables}
\label{sec:bayes}

Suppose we are given some data observations $\mathcal{D}$, and let $\mathcal{D}_S = \{x_S^{(i)}, y_S^{(i)}\}^{N_S}_{i=1}$ indicate the source domain data with input $x_S$ and label $y_S$, and $\mathcal{D}_T = \{x_T^{(i)}, y_T^{(i)}\}^{N_T}_{i=1}$ indicate the target domain data with input $x_T$ and label $y_T$, respectively.
In knowledge transfer, the amount of source data is much larger than the amount of target data, i.e., ($N_S \gg N_T$), which implies low-resource conditions in the target domain.

Consider now a DNN-based discriminative model with parameters $\lambda$ to be estimated, where $\lambda$ represents the network weights. Starting from the classical Bayesian approach, a prior distribution $p(\lambda)$ is defined over $\lambda$. We hereby define the posterior distribution of $\lambda$ as $p(\lambda|\mathcal{D})$, and the likelihood of data $\mathcal{D}$ as $p(\mathcal{D})$. After observing the data $\mathcal{D}$, $p(\lambda|\mathcal{D})$ can be obtained by the Bayes Rule:
\vspace{0.03cm}
\begin{equation}
    p(\lambda|\mathcal{D}) = \frac{p(\mathcal{D}|\lambda) p(\lambda)}{p(\mathcal{D})}.
    \label{eq:bayes}
\end{equation}

In addition to the network weights, we introduce latent variables $Z$ to model the intermediate DNN hidden embedding. Here $Z$ refers to the unobserved intermediate representations, which usually encode transferable distributional information and structural relationships among data. Figure \ref{fig:framework} illustrates the proposed overall framework.
Specifically, $Z$ can correspond to the hidden embedding of either intermediate or final outputs of a deep model.
We can decouple the network weights into two independent subsets, $\theta$ and $\omega$ to represent weights before and after $Z$ is generated, respectively, as illustrated by the four networks in Figure \ref{fig:framework}. Typically, $\theta$ encompasses the majority of the model’s parameters, allowing $Z$ to capture rich, transferable feature representations.
Considering the entire deep neural network, we have:
\vspace{0.03cm}
\begin{equation}
    p(\lambda) = p(Z, \theta, \omega) = p(Z|\theta) p(\theta) p(\omega).
    \label{eq:lambda}
\end{equation}

The relationship in Eq. (\ref{eq:lambda}) holds for both the prior $p(\lambda)$and the posterior $p(\lambda|\mathcal{D})$.
For a special case, when $Z$ represents the logits (model outputs before softmax) or soft output (model outputs after softmax), the mapping network from $Z$ to $Y$ has no-trainable parameters, since $\omega$ is a constant, and $p(\omega)$ is non-informative.

In this work, we focus on transferring knowledge in a distributional sense via the latent variables $Z$.
We denote $\{Z, \theta, \omega\}$ for source and target domains as $\{Z_S, \theta_S, \omega_S\}$ and $\{Z_T, \theta_T, \omega_T\}$, respectively. We also assume that the deep latent variables, $Z$, retain the same distributions across source and target domains. Specifically, for the target latent variables, the prior knowledge learned from the source domain is:
\vspace{0.03cm}
\begin{equation}
    p(Z_T|\theta_T) = p(Z_S|\theta_S, \mathcal{D}_S).
    \label{eq:prior}
\end{equation}

The target posterior distribution, namely $p(\lambda_T|\mathcal{D}_T)$, is often intractable, and an approximation is required. There exist several techniques to obtain such an approximation and solve Eq. (\ref{eq:bayes}). For instance, maximum a posteriori (MAP) \cite{degroot2005optimal, duda1973pattern}, variation Bayes (VB) \cite{beal2003variational, graves2011practical}, and Markov Chain Monte Carlo (MCMC) \cite{neal1993probabilistic}. In this work, we consider the VB-based framework and obtain a full approximation of the posterior distribution.

In knowledge transfer settings, parallel data is usually utilized to obtain an effective knowledge transfer process \cite{lopez2015unifying, vapnik2015learning, li2017large}. The existence of parallel data indicates that there exists a paired source data sample $x_S^{(j)}$  for each target data sample $x_T^{(i)}$. Therefore, $x_T^{(i)}$ and $x_S^{(j)}$ have strong relationships: same sample except for the domain information, e.g., the same audio content but recorded by different devices. We thus refer to the paired $x_T^{(i)}$ and $x_S^{(j)}$ as parallel data samples for the knowledge transfer task. Such data can provide strong domain information and allow the target model to learn prior knowledge from the source domain. In Section \ref{sec:vbkt}, we will describe how to take advantage of parallel data and how to deal with its absence.

\vspace{0.2cm}
\section{Variational Bayesian Adaptive Learning}
\label{sec:vbkt}

\subsection{Variational Bayesian Adaptation}

To approximate the posterior density $p(\lambda_T|\mathcal{D}_T)$ within a VB framework, we need to introduce a variational distribution $q(\lambda_T|\mathcal{D}_T)$. In the target domain, the optimal solution, $\hat{q}(\lambda_T|\mathcal{D}_T)$, is obtained by minimizing the KL divergence (KLD, \cite{kullback1951information}) between $q(\lambda_T|\mathcal{D}_T)$ and the real one $p(\lambda_T|\mathcal{D}_T)$, over a family of allowed approximate distributions $\mathcal{Q}$:
\vspace{0.03cm}
\begin{align}
    &\hat{q}(\lambda_T|\mathcal{D}_T) \nonumber \\
    & \quad\quad = \mathop{argmin}\limits_{q\in \mathcal{Q}}\  \mathtt{KLD}(q(\lambda_T|\mathcal{D}_T) \ \|\  p(\lambda_T|\mathcal{D}_T))\\
    & \quad\quad = \mathop{argmin}\limits_{q\in \mathcal{Q}}\  \log p(\mathcal{D}_T) - \mathds{E}_{\lambda_T \sim q(\lambda_T| \mathcal{D}_T)} \log p(\mathcal{D}_T | \lambda_T) \nonumber\\
    & \quad\quad\quad\quad\quad\quad\quad\quad + \mathtt{KLD} (q(\lambda_T|\mathcal{D}_T)\ \|\ p(\lambda_T)).
    \label{eq:approx}
\end{align}
\vspace{0.03cm}

In Eq. (\ref{eq:approx}), $p(\mathcal{D}_T)$ represents the natural distribution of the target data; therefore, we can simply omit it during the approximation. In this work, we consider the latent variables $z$ and assume a non-informative prior (e.g., a uniform distribution) over the model weights, specifically \( \theta_T \) and \( \omega_T \). By substituting Eq. (\ref{eq:lambda}) into Eq. (\ref{eq:approx}) and rearranging the terms, we derive the following evidence lower bound (ELBO, also known as the variational lower bound) $\mathcal{L}(\lambda_T; \mathcal{D}_T)$:
\vspace{0.03cm}
\begin{align}
    \mathcal{L}(\lambda_T; \mathcal{D}_T) &= \mathds{E}_{Z_T \sim q(Z_T| \theta_T, \mathcal{D}_T)} \log p(\mathcal{D}_T |Z_T, \theta_T, \omega_T) \nonumber\\
     &\quad - \mathtt{KLD} (q(Z_T|\theta_T, \mathcal{D}_T)\ \|\ p(Z_T|\theta_T))
    \label{eq:elbo_0}\\
    &= \mathds{E}_{Z_T \sim q(Z_T| \theta_T, \mathcal{D}_T)} \log p(\mathcal{D}_T |Z_T, \theta_T, \omega_T) \nonumber\\
     &\quad - \mathtt{KLD} (q(Z_T|\theta_T, \mathcal{D}_T)\ \|\ p(Z_S|\theta_S, \mathcal{D}_S)).
    \label{eq:elbo}
\end{align}
\vspace{0.03cm}

Eq. (\ref{eq:elbo}) is derived by substituting Eq. (\ref{eq:prior}) into Eq. (\ref{eq:elbo_0}), which facilitates the knowledge transfer process. Consequently, with Eq. (\ref{eq:elbo}) serving as the optimization objective, we propose and develop a variational Bayesian adaptation approach.

To solve the ELBO in Eq. (\ref{eq:elbo}), it is necessary to specify the distributional forms for both the prior and posterior probabilities over $Z$ to achieve a closed-form solution for the second term in Eq. (\ref{eq:elbo}). In this work, we employ two different approximations to address the presence or absence of parallel data: (i) Gaussian mean-field variational inference (GMFVI) \cite{mezard1987spin, ambjorn1997quantum, kingma2013auto} is utilized when parallel data is available, and (ii) empirical Bayes (EB), also known as evidence approximation \cite{bishop2006pattern, robbins2020empirical}, is applied when parallel data is not available.

\subsection{Estimation with GMFVI}
\label{sec:gmfvi}

When parallel data is available, a Gaussian mean-field approximation is used to specify the distribution forms for both the prior and posterior over $Z$. In this context, the variational Bayesian knowledge transfer method is referred to as VBKT-GMF. The latent variables are defined as $Z = \{Z^{(i)}\}_{i=1}^{N_T}$ where $N_T$ represents the number of target training samples. If $M$ is the hidden embedding size, each $Z^{(i)}$ follows an $M$-dimensional Gaussian distribution. We assume that all Gaussian components $Z^{(i, j)}$ within the multivariate Gaussian  $Z^{(i)}$ are independent of each other, resulting in a diagonal covariance matrix. Therefore, $Z^{(i, j)} \sim \mathcal{N}(Z^{(i, j)}; \mu^{(i, j)}, (\sigma^{(i, j)})^2)$. Finally, the variational distribution  $q(Z | \theta, \mathcal{D})$ can be expressed as:
\vspace{0.03cm}
\begin{align}
    q(Z | \theta, \mathcal{D}) &= \prod^{N_T}_i \mathcal{N}(Z^{(i)}; \mu^{(i)}, (\sigma^{(i)})^2 \bm{\mathcal{I}}) \nonumber \\
    &= \prod^{N_T}_i \prod^{M}_j \mathcal{N}(Z^{(i, j)}; \mu^{(i, j)}, (\sigma^{(i, j)})^2).
    \label{eq:gaussians}
\end{align}

Given a parallel data set, each $Z^{(i)}$ corresponds to an individual input sample $x^{(i)}$, and each source domain data sample $x^{(i)}_S$ pairs with a target domain sample $x^{(i)}_T$. We can approximate the KL divergence term in Eq. (\ref{eq:elbo}) by establishing a mapping of each pair of Gaussian distributions across domains via sample pairs. Denoting the Gaussian mean and variance for the source and target domains as ($\mu_S, \sigma_S^2$), and ($\mu_T, \sigma_T^2$), respectively, we can obtain a closed-form solution for the KLD term by substituting Eq. (\ref{eq:gaussians}) into Eq. (\ref{eq:elbo}):
\vspace{0.03cm}
\begin{align}
    &\mathtt{KLD} (q(Z_T|\theta_T, \mathcal{D}_T)\ \|\ p(Z_S|\theta_S, \mathcal{D}_S)) \nonumber\\
    &\quad =\sum_i^{N_T} \sum_j^{M} (\log \frac{\sigma_S^{(i, j)}}{\sigma_T^{(i, j)}} + \frac{(\sigma_T^{(i, j)})^2 + (\mu_T^{(i, j)} - \mu_S^{(i, j)})^2}{2 (\sigma_S^{(i, j)})^2} - \frac{1}{2}).
    \label{eq:kld_0}
\end{align}

A stochastic gradient variational Bayesian (SGVB) estimator \cite{kingma2013auto} can be used to approximate the posterior, with the mean of the Gaussian generated by the network's hidden outputs, determined by the weights $\theta$ and input $x^{(i)}$. For the variance ($\sigma^{(i)})^2$, it can either be approximated with a network or considered as a trainable variable. In this work, we assign a fixed value $\sigma^2$ to both $(\sigma_S^{(i)})^2$ and $(\sigma_T^{(i)})^2$. We can now rewrite Eq. (\ref{eq:kld_0}) as follows:
\vspace{0.03cm}
\begin{align}
    \mathtt{KLD} (q(Z_T|\theta_T, \mathcal{D}_T)\ &\|\ p(Z_S|\theta_S, \mathcal{D}_S)) \nonumber\\
    & = \frac{1}{2\sigma^2} \sum_i^{N_T} \| \mu_T^{(i)} - \mu_S^{(i)}\|_2^2.
    \label{eq:kld}
\end{align}

Furthermore, by adopting Monte Carlo method to generate $N$-pairs of sample, the lower bound in Eq. (\ref{eq:elbo}) can be approximated empirically approximated as follows:
\vspace{0.03cm}
\begin{align}
    &\mathcal{L}(\lambda_T; \mathcal{D}_T) \nonumber\\
    &\quad\quad = \sum^{N_T}_i \mathds{E}_{z^{(i)}_T \sim \mathcal{N}(\mu_T^{(i)}, \sigma^2)} \log p(y_T^{(i)} | x_T^{(i)}, z^{(i)}_T, \theta_T, \omega_T) \nonumber\\
    &\quad\quad\quad\quad - \frac{1}{2\sigma^2} \sum_i^{N_T} \| \mu_T^{(i)} - \mu_S^{(i)}\|_2^2,
    \label{eq:elbo_new}
\end{align}
where the first term in RHS represents the likelihood, and the second term is obtained from the KL divergence between the prior and posterior of the latent variables. Each instance $z^{(i)}_T$ of $Z^{(i)}_T$, is sampled from the variational posterior distribution $q(Z_T^{(i)}|\theta_T, \mathcal{D}_T)$ so that the expectation form in the first term can be reduced.
Moreover, it should be noted the second term differs from a simple $L_2$ distance between hidden features since the measurement is carried out over a probabilistic distribution space.
In the training stage, the gradient of the sampling operation flows through the layers of the deep nets thanks to a reparameterization trick \cite{salimans2013fixed, kingma2013auto}.
In the inference stage, computational complexity is simplified by using $\mu^{(i)}_T$ for $z^{(i)}_T$.

\subsection{Estimation with EB}
\label{sec:eb}

\begin{figure}[t]
  \centering
  \vspace{0.2cm}
  \includegraphics[width=0.86\linewidth]{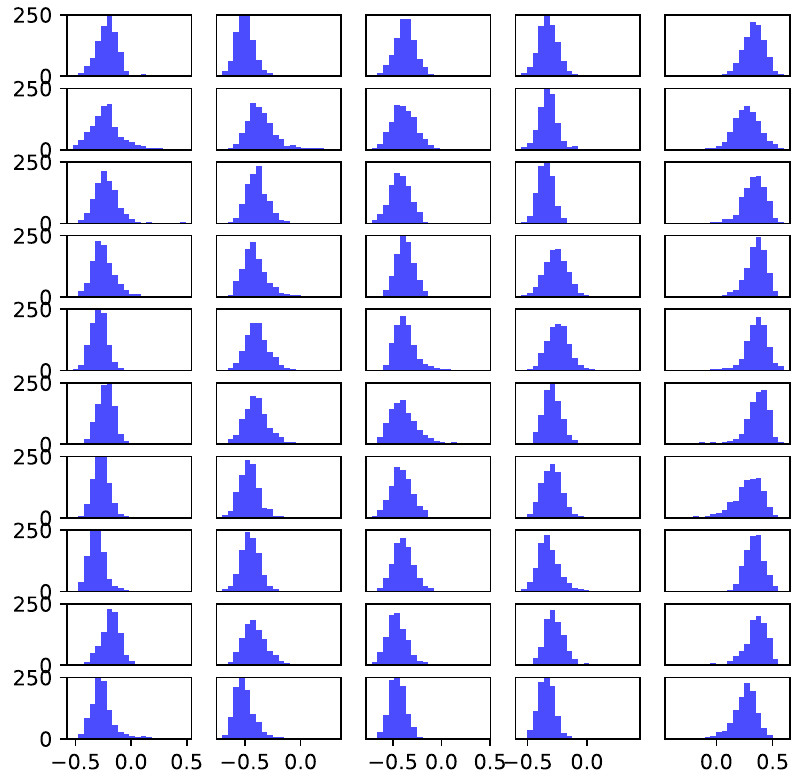}
  \vspace{0.3cm}
  \caption{Histograms of sample hidden embedding elements from 5 classes. This visualization is performed using the CRNN-att model of the SCR task. Each subplot represents an element of the hidden embedding, as each one is accumulated by class index, where subplots in the same row correspond to the same class.}
  \label{fig:gaussians}
\end{figure}

In the GMFVI approach, the prior knowledge learned from the source domain is encoded in the distribution of latent variables, $Z$, and then transferred to the target domain via parallel data samples. However, a mean-field solution is difficult to put forth if parallel data is not available. In real-world cross-domain knowledge transfer, parallel data is rarely available, so a solution needs to be devised. To overcome such a problem, we leverage empirical Bayes \cite{degroot1970optimal} and denote the whole approach as VBKT-EB.

To perform VBKT-EB, a prior distribution over the deep latent variables must be imposed. We start by considering the source domain data as history observations to estimate the hyperparameters of the prior distributions, employing an empirical Bayes approach. Figure \ref{fig:gaussians} shows histograms of the latent variable $Z$, with each subplot representing an element of the hidden embedding, specifically the $i$-th element of an $M$-dimensional hidden feature vector. Features are accumulated by class index, where subplots in the same row correspond to the same class. It can be observed that the distribution of the hidden embedding for each class exhibits the characteristics of a Gaussian distribution. This pattern holds true for other elements and classes as well. Therefore, we assume that the distribution of latent variables $Z$ is a mixture of joint Gaussian distributions.
To be specific, suppose that there are $C$ classes, the hidden embedding for the same class is assumed to belong to a multi-variant Gaussian distribution. Therefore, we have $C$ components for the mixture of Gaussian, as $Z = \{Z^{(c)}\}_{c=1}^{C}$. We assume the covariance matrix to be diagonal. Thus, we have $C$ multi-variant Gaussian distributions to represent the prior, as $Z^{(c)} \sim \mathcal{N}(\mu^{(c)}, (\sigma^{(c)})^2 \bm{\mathcal{I}})$, where $\bm{\mathcal{I}}$ is an identity matrix. Leveraging upon Eq. (\ref{eq:prior}), for each $Z_T^{(c)}$ in $Z_T$, the prior density distribution is:
\begin{align}
    p(Z^{(c)}_T|\theta_T) = p(Z^{(c)}_S|\theta_S, \mathcal{D_S}) = \mathcal{N}(Z^{(c)}_S; \mu^{(c)}, (\sigma^{(c)})^2 \bm{\mathcal{I}}).
    \label{eq:prior_eb}
\end{align}

Given $N_S^{(c)}$ source domain observations of the latent variable $Z_S^{(c)}$, we can perform maximum likelihood estimation (MLE) to determine the mean $\mu^{(c)}$ and variance $(\sigma^{(c)})^2$ of the Gaussian distribution. These parameters can be expressed as:
\begin{align}
    \mu_{MLE}^{(c)} &= \frac{1}{N_S^{(c)}} \sum_i^{N_S^{(c)}} z_S^{(c,i)}, \label{eq:eb_mean}\\
    (\sigma_{MLE}^{(c)})^2 &= \frac{1}{N_S^{(c)}} \sum_i^{N_S^{(c)}} (z_S^{(c, i)} - \mu_{MLE}^{(c)})(z_S^{(c, i)} - \mu_{MLE}^{(c)})^T, \label{eq:eb_var}
\end{align}
where $z_S^{(c, i)}$ indicates the hidden embedding generated by the input $x_S^{(c, i)}$ and  parameters $\theta_S$, from the class $c$.

For the target variational posterior, $q(Z_T|\theta_T, \mathcal{D}_T)$, we denote the Gaussian mean and variance with $\mu_T$, and $\sigma_T^2$, respectively. Given the EB-estimated prior density in Eqs. (\ref{eq:eb_mean}) and (\ref{eq:eb_var}), a closed-form solution for KLD in Eq. (\ref{eq:elbo}) can be obtained as follows:
\begin{align}
    &\mathtt{KLD} (q(Z_T|\theta_T, \mathcal{D}_T)\ \|\  p(Z_S|\theta_S, \mathcal{D}_S)) \nonumber\\
    &\quad\quad = \sum_c^{C} \sum_i^{N_T^{(c)}} \sum_j^{M} (\log \frac{\sigma_{MLE}^{(c, j)}}{\sigma_T^{(c, i, j)}} \nonumber\\
    &\quad\quad\quad\quad + \frac{(\sigma_T^{(c, i, j)})^2 + (\mu_T^{(c, i, j)} - \mu_{MLE}^{(c, j)})^2}{2 (\sigma_{MLE}^{(c, j)})^2} - \frac{1}{2}),
    \label{eq:kld_eb}
\end{align}
where $C$, $N_T^{(c)}$, $M$ represents the number of target classes, the number of target training samples with the label of class $c$, and the size of hidden embedding, respectively.

Similar to the VBKT-GMF case, a stochastic gradient variational Bayesian estimator \cite{kingma2013auto} is used to approximate the posterior distribution. The target domain Gaussian mean, $\mu_T^{(c, i)}$, is generated by the network hidden outputs and determined by the weights $\theta$ and input $x^{(c, i)}$ for class $c$.
Moreover, $(\sigma_T^{(c, i)})^2$ is set equal to $(\sigma_{MLE}^{(c)})^2$ for all the target input samples, $x_T^{(c, i)}$, from class $c$. Re-arranging the terms and adopting Monte Carlo sampling in  Eq. (\ref{eq:kld_eb}) allows us to  empirically approximate the ELBO in Eq. (\ref{eq:elbo}) as follows:
\begin{align}
    &\mathcal{L}(\lambda_T; \mathcal{D}_T) \nonumber\\
    &\quad\quad = \sum^{N_T}_i \mathds{E}_{z^{(i)}_T \sim \mathcal{N}(\mu_T^{(i)}, (\sigma_{MLE}^{(c)})^2)} \log p(y_T^{(i)} | x_T^{(i)}, z^{(i)}_T, \theta_T, \omega_T) \nonumber\\
    &\quad\quad\quad\quad - \frac{1}{2(\sigma_{MLE}^{(c)})^2} \sum_c^C \sum_i^{N^{(c)}_T} \| \mu_T^{(c, i)} - \mu_{MLE}^{(c)}\|_2^2.
    \label{eq:elbo_eb}
\end{align}

The prior knowledge is encoded in density $p(Z_S|\theta_S, \mathcal{D_S})$, obtained by the EB method from the source domain observations, and then transferred to a target domain by optimizing Eq. (\ref{eq:elbo_eb}). Note that parallel data is not required, since the hyper-parameters $\mu_{MLE}^{(c)}$ and $(\sigma_{MLE}^{(c)})^2$ are not related to the target domain data. The proposed EB approach also employs a reparameterization trick \cite{salimans2013fixed, kingma2013auto} to allow the gradient to flow through the sampling operation in the neural model.

\subsection{Structural Relationship Modeling}
\label{sec:struct}

For both GMFVI and EB estimation approaches, a Gaussian distribution is employed to (i) approximate the variational distribution and (ii) estimate the KL divergence between the target posterior and the prior learned from the source domain (see the second term on the RHS in Eq. (\ref{eq:elbo})). However, for these two estimation methods, the Gaussian components are assumed to be independent of one another. The same number of Gaussian components is used to model latent variables in both the source and target models, with a Gaussian-to-Gaussian matching across domains. Consequently, we approximate the KL divergence for each Gaussian pair from the two domains, disregarding the structural relationships among deep latent variables during adaptation.

However, as highlighted by various authors (e.g., \cite{hu2020relational, park2019relational, peng2019correlation}), latent variables within deep neural networks usually encapsulate structural relationships. Incorporating the interdependencies among Gaussian distributions can significantly improve the approximation of the KL divergence between Gaussian mixtures \cite{hershey2007approximating}. Therefore, we propose introducing structural relationship modeling as a means to enhance the approximation of the KL divergence between two Gaussian mixtures. Our focus lies on modeling the structural relationships among each pair of Gaussian components within the same mixture. The KL divergence serves as the distance measure between two Gaussian components. Given $N_G$ distribution components in a mixture $p(Z)$, where $Z={Z^{(i)}}_{i=1}^{N_G}$, we can formulate the overall intra-mixture relational distance value $\mathcal{V} (p(Z))$ as:
\vspace{0.03cm}
\begin{align}
    \mathcal{V} (p(Z)) &= \frac{1}{N_G^2} \sum_i^{N_G} \sum_j^{N_G} \mathtt{KLD} ( p(Z^{(i)})\ \|\  p(Z^{(j)})).
    \label{eq:rela_val_0}
\end{align}
\vspace{0.03cm}

In the same spirit of the GMFVI and EB approaches, we can consider $p(Z)$ to be a mixture of $M$-dimension Gaussian distributions with a diagonal covariance matrix, where $Z^{(i)} \sim \mathcal{N} (Z^{(i)}; \mu^{(i)}, \sigma^{(i)}\bm{\mathcal{I}})$, and   Eq. (\ref{eq:rela_val_0}) becomes:
\begin{align}
    \mathcal{V} (p(Z)) &= \frac{1}{N_G^2} \sum_i^{N_G} \sum_j^{N_G} \sum_m^{M} (\log \frac{\sigma^{(j, m)}}{\sigma^{(i, m)}} \nonumber\\
    &\quad\quad + \frac{(\sigma^{(i, m)})^2 + (\mu^{(i, m)} - \mu^{(j, m)})^2}{2(\sigma^{(j, m)})^2} - \frac{1}{2}).
    \label{eq:rela_val}
\end{align}
\vspace{0.03cm}

Then the new criterion is built by minimizing the Huber loss \cite{huber1992robust} of the overall relational distance from two mixtures. The Huber loss (also named as smoothed $L_1$ loss) is given by
\begin{align}
SL1 (x, y)=\left\{
\begin{aligned}
\frac{1}{2} (x - y)^2, |x - y| \leq 1, \\
|x - y| - \frac{1}{2}, |x - y| > 1.
\end{aligned}
\right.
\label{eq:huber}
\end{align}
\vspace{0.03cm}

The introduction of relationship modeling allows us to  re-formulate the ELBO in Eq. (\ref{eq:elbo}) as follows:
\begin{align}
    &\mathcal{L}(\lambda_T; \mathcal{D}_T) \nonumber\\
    &\quad = \mathds{E}_{Z_T \sim q(Z_T| \theta_T, \mathcal{D}_T)} \log p(\mathcal{D}_T |Z_T, \theta_T, \omega_T) \nonumber\\
    &\quad\quad\quad - \mathtt{KLD} (q(Z_T|\theta_T, \mathcal{D}_T)\ \|\ p(Z_S|\theta_S, \mathcal{D}_S)) \nonumber\\
     &\quad\quad\quad - \beta * SL1 (\mathcal{V}(q(Z_T|\theta_T, \mathcal{D}_T)), \mathcal{V}(p(Z_S|\theta_S, \mathcal{D}_S))),
     \label{eq:elbo_rela}
\end{align}
where $\beta$ is a tunable parameter to balance the loss terms.
We denote VBKT using the new objective in \ref{eq:elbo_rela} as VBKT-GMF-rela and VBKT-EB-rela, with respect to the GMFVI and EB estimation ways, respectively.
Generally, the original objective in Eq. (\ref{eq:elbo}) can well approximate the KL divergence between two Gaussian mixtures in most cases. The plug-in relationship term (the third term in RHS of Eq. (\ref{eq:elbo_rela})) is likely to enhance the approximation, but it may also play an important role when structural information between samples or classes is important for the adaptation.

\section{Experiments}
\label{sec:exp}

\subsection{Datasets \& Baseline Systems}

We assess our proposed VBKT methodologies using two audio-based data sets.
The first is the DCASE 2020 challenge set, focusing on acoustic scene classification (ASC) \cite{heittola2020acoustic}
to assess the proposed VBKT with GMFVI prior estimation since it provides parallel source-target materials. 
The data set contains recordings of ten different acoustic scenes from multiple European cities.
The training set comprises $\sim$10K scene audio clips recorded by the source device (device A), and 750 clips for each of the eight target devices (Device B, C, s1-s6).
Audio recordings collected with devices A, B, and C are real-world data and recorded simultaneously. Devices s1-s6 are additional synthetic data simulated leveraging audio clips recorded with device A, which is a high-quality binaural device.
Simulated recordings are created with a dataset of impulse responses (IR) measured for multiple angles using mobile devices. 

Each target audio clip is paired with a source audio clip, and the recording device is the only difference between the two audio clips.
The objective is to address the device mismatch problem for each target device individually, i.e., device adaptation, which is a prevalent challenge in real-world applications.
For each audio clip, log-mel filter bank (LMFB) \cite{davis1980comparison} features are extracted and scaled to the range [0,1] before feeding into the classifier.
Three state-of-the-art models, namely: a dual-path residual convolutional neural network (RESNET) \cite{hu2021two}, a fully convolutional neural network with channel attention (FCNN) \cite{hu2021two}, and a low-complexity shallow inception model (INCEPTION) \cite{yen2021lottery} are tested according to the DCASE challenge outcomes \cite{heittola2020acoustic, martin2021low}.
The INCEPTION model is designed for low-complexity on-device applications so it features significantly fewer network parameters.
The same models are used for both source and target devices. During the training phase, we implement Mix-up \cite{zhang2017mixup} and SpecAugment \cite{park2019specaugment} techniques.
The mix-up alpha parameter is set to 0.4 and applied at the mini-batch level: where two data batches along with their respective labels are randomly mixed during each training step.
As for the SpecAugment, each feature map is randomly masked along both time and frequency axes, with 10\% of the dimensions set to zero.

The second data set is a limited-vocabulary yet notably challenging speech command recognition (SCR) \cite{warden2018speech} task, namely the Google Speech Command Dataset v2.
For SCR, we employ a ten-class setting from earlier studies \cite{warden2018speech, de2018neural, yang2021decentralizing},
incorporating a total of $\sim$30K training utterances as source domain data.
We identify four target domains, each comprising 400 utterances mixed with four distinct types of noise from the data set, labeled N1 through N4. 
The key difference between the source and target domains is the background noise.
All types of noise are added based on different signal noise ratio (SNR) values, namely [10, 15, 20, 25, 30]. Target utterances have no overlaps with source utterances, thus parallel material is not available.
We adopt a benchmark deep attention convolutional recurrent neural network model (CRNN-att)  \cite{de2018neural} as a baseline for both source and target models, since it has attained a state-of-the-art performance \cite{zhang2017hello, jansson2018single, yen23_interspeech}.

\begin{table}[t]
\centering
\vspace{0.2cm}
\caption{13 cutting-edge knowledge transfer methods, and their tunable parameter settings, compared in our experiments. $\alpha$ and $\beta$ indicate the parameters to modulate the weight of the introduced adaptation loss, whereas $T$ is the temperature parameter used for softmax computation.}
\label{tab:para_setting}
\small{
\begin{tabular}{l|l}
\toprule
\toprule
Baseline system        &  Settings           \\
\midrule
\midrule
Teacher-Student Learing (TSL)   \cite{hinton2015distilling}        & $T$=1.0                              \\
Neural Label Embedding (NLE)    \cite{meng2020vector}       & S-KLD, $T$=1.0                \\
Hints for Thin Nets (Fitnets)  \cite{romero2014fitnets}     & $\alpha$=2.0                         \\
Attention Transfer (AT)     \cite{zagoruyko2016paying}       & $\alpha$=1000                        \\
Activation Boundaries (AB)    \cite{heo2019knowledge}        & $\alpha$=0 \\
Variational Information Distillation (VID)    \cite{ahn2019variational}       & $\alpha$=1.0                         \\
Flow of Solution Procedure (FSP)   \cite{yim2017gift}        & $\alpha$=0 \\
Comprehensive Overhaul (COFD)    \cite{heo2019comprehensive}       & $\alpha$=0.5                         \\
Similarity Preserving (SP)   \cite{tung2019similarity}          & $\alpha$=300                         \\
Correlation Congruence (CCKD)   \cite{peng2019correlation}       & $\alpha$=0.02                        \\
Probabilistic Knowledge Transfer (PKT)   \cite{passalis2018learning}        & $\alpha$=30000                       \\
Neuron Selectivity Transfer (NST)  \cite{huang2017like}         & $\alpha$=10.0                        \\
Relational Knowledge Distillation (RKD)   \cite{park2019relational}        & $\alpha$=25, $\beta$=50              \\
\bottomrule
\bottomrule
\end{tabular}
}
\end{table}

\subsection{Experimental Setup}

The latent variables for each task are based on the model's hidden layer before the last layer outputs.
For instance, within the ASC task, we utilize the hidden embedding after batch normalization but before ReLU activation from the second last convolutional layer for the RESNET, FCNN, and INCEPTION models.
As for the CRNN-att model used in the SCR task, the hidden embedding before the final activation layer is utilized.
As previously mentioned, in the variational inference step to approximate the posterior distribution, a deterministic value is set for the standard deviation $\sigma_T$ of Gaussians for simplicity. This deterministic value is obtained from the augmented data.
Specifically, we employ data augmentation algorithms recommended in \cite{hu2021two} to generate additional data for each audio clip.
Thus each audio clip possesses a unique set of augmented data based upon itself. Then the average standard deviation over each set is computed.
As for the structural relationship modeling, $\beta = 0.1$ is used.

An extensive experimental comparison, against thirteen cutting-edge knowledge transfer techniques listed in Table \ref{tab:para_setting}, is carried out. We adhere to the recommended setups and parameter settings outlined in the original papers.
Tunable parameter settings for all methods are listed in Table \ref{tab:para_setting}.
The majority of methods featured in Table \ref{tab:para_setting} introduce an additional adaptation loss and utilize a parameter $\alpha$ to modulate the weight of this adaptation loss.
In doing so, the final loss is formed as a linear combination of the original classification loss (e.g., cross-entropy) and the adaptation loss. As for the AB \cite{heo2019knowledge} and FSP \cite{yim2017gift} methods, the authors recommend pre-training and fine-tuning procedures. The latter implies that the adaptation technique is used in the pre-training stage only; whereas $\alpha = 0$ in the fine-tuning stage.

\begin{table}[t]
\vspace{0.1cm}
\centering
\caption{Evaluation accuracies (in \%) of the source model on 9 devices of the DCASE2020 ASC data set. Device A is the source device and others (including 8 devices: B, C, s1-s6) are target devices.}
\label{tab:res-asc-base}
\begin{tabular}{l||ccc}
\toprule
\toprule
                  & RESNET & FCNN & INCEPTION                 \\
\midrule
\midrule
Device A  (\%)       & 79.09                      & 79.70                     & 78.52 \\
8 other devices (\%) & 37.70                      & 37.13                    &           32.35               \\
\bottomrule
\bottomrule
\end{tabular}
\end{table}

\begin{table*}[t]
\vspace{0.2cm}
\centering
\caption{A comparison of average evaluation accuracies (in \%) on recordings of the DCASE2020 ASC data set. The proposed method VBKT with GMFVI is evaluated. Each cell represents the average value over 32 experimental results for 8 target devices $\times$ 4 repeated trails. The standard deviation value is computed on 4 averages over 8 target devices. TSL represents the teacher-student learning.}
\label{tab:res-asc}
\begin{tabular}{c|l||c|c|c||c|c|c}
\toprule
\toprule
\multicolumn{2}{l||}{\multirow{2}{*}{\ System}}                     & \begin{tabular}[c]{@{}c@{}}RESNET\\ avg\% $\pm$ std\end{tabular}                                 & \begin{tabular}[c]{@{}c@{}}FCNN\\ avg\% $\pm$   std\end{tabular}                                     & \begin{tabular}[c]{@{}c@{}}INCEPTION\\ avg\% $\pm$ std\end{tabular}  & \begin{tabular}[c]{@{}c@{}}RESNET\\ avg\% $\pm$ std\end{tabular}                                    & \begin{tabular}[c]{@{}c@{}}FCNN\\ avg\% $\pm$  std\end{tabular}                                  & \begin{tabular}[c]{@{}c@{}}INCEPTION\\ avg\% $\pm$  std\end{tabular}  \\
\cmidrule{3-8}
\multicolumn{2}{l||}{}                   &
                                                             \multicolumn{3}{c||}{\textit{without} TSL}                                                                         & \multicolumn{3}{c}{\textit{with} TSL}                                                                          \\
\midrule
\midrule
\ (1) & Source model                                     & 37.70                                       & 37.13                                      &     32.35      & -                                          & -                                          & -         \\
\ (2) &  No-transfer                                   & 54.29 $\pm$ {\small 0.76}                           & 49.97 $\pm$ {\small 2.70}                           &   52.25 $\pm$ {\small 1.31}       & -                                          & -                                          & -         \\
\ (3) &  One-hot                             & 63.76 $\pm$ {\small 0.51}                           & 64.45 $\pm$ {\small 0.51}                           &    64.38 $\pm$ {\small 0.58}      & -                                          & -                                          & -         \\
\midrule
\ (4) &  TSL                                           & 68.04 $\pm$ {\small 0.34}                           & 66.27 $\pm$ {\small 0.46}                           &    65.59 $\pm$ {\small 0.52}      & 68.04 $\pm$ {\small 0.34}                           & 66.27 $\pm$ {\small 0.46}                           &      65.59 $\pm$ {\small 0.52}    \\
\ (5) &  NLE                                           & 65.64 $\pm$ {\small 0.53}                           & 64.47 $\pm$ {\small 0.59}                           &    63.17 $\pm$ {\small 0.35}      & 67.76 $\pm$ {\small 0.63}                           & 64.53 $\pm$ {\small 0.56}                           &    63.80 $\pm$ {\small 0.29}      \\
\ (6) &  Fitnets                                        & 66.73 $\pm$ {\small 0.34}                           & 67.29 $\pm$ {\small 0.39}                           &    65.80 $\pm$ {\small 0.16}      & 69.89 $\pm$ {\small 0.21}                          & 69.06 $\pm$ {\small 0.30}                           &   67.88 $\pm$ {\small 0.19}       \\
\ (7) &  AT                                            & 63.73 $\pm$ {\small 0.81}                           & 64.16 $\pm$ {\small 0.49}                          &    64.25 $\pm$ {\small 0.46}      & 68.06 $\pm$ {\small 0.45}                         & 66.35 $\pm$ {\small 0.52}                           &     65.98 $\pm$ {\small 0.17}     \\
\ (8) &  AB                                            & 65.34 $\pm$ {\small 0.54}                           & 66.21 $\pm$ {\small 0.59}                          &   64.58 $\pm$ {\small 0.89}       & 68.69 $\pm$ {\small 0.47}                           & 66.91 $\pm$ {\small 0.17}                          &      66.73 $\pm$ {\small 0.28}    \\
\ (9) &  VID                                           & 63.90 $\pm$ {\small 0.67}                           & 63.79 $\pm$ {\small 0.37}                          &   64.27 $\pm$ {\small 0.38}       & 68.56 $\pm$ {\small 0.49}                           & 65.75 $\pm$ {\small 0.40}                           &     65.89 $\pm$ {\small 0.39}     \\
\ (10) &  FSP                                           & 64.44 $\pm$ {\small 0.65}                           & 65.33 $\pm$ {\small 0.64}                           &  64.46 $\pm$ {\small 0.62}        & 68.94 $\pm$ {\small 0.32}                           & 66.01 $\pm$ {\small 0.15}                           &     66.36 $\pm$ {\small 0.51}     \\
\ (11) &  COFD                                          & 64.92 $\pm$ {\small 0.81}                           & 66.69 $\pm$ {\small 0.23}                           &  64.45 $\pm$ {\small 0.25}        & 68.57 $\pm$ {\small 0.34}                           & 68.63 $\pm$ {\small 0.34}                           &     66.01 $\pm$ {\small 0.47}     \\
\ (12) &  SP                                            & 64.57 $\pm$ {\small 0.76}                           & 65.74 $\pm$ {\small 0.37}                           &  65.10 $\pm$ {\small 0.39}        & 68.45 $\pm$ {\small 0.28}                           & 67.36 $\pm$ {\small 0.33}                           &    66.40 $\pm$ {\small 0.22}      \\
\ (13) &  CCKD                                          & 65.59 $\pm$ {\small 0.47}                           & 66.52 $\pm$ {\small 0.30}                           &  63.73 $\pm$ {\small 0.90}        & 69.47 $\pm$ {\small 0.40}                           & 68.29 $\pm$ {\small 0.25}                           &     64.45 $\pm$ {\small 0.36}     \\
\ (14) &  PKT                                           & 64.65 $\pm$ {\small 0.68}                           & 64.84 $\pm$ {\small 0.24}                           &   65.22 $\pm$ {\small 0.29}       & 65.43 $\pm$ {\small 0.87}                           & 67.25 $\pm$ {\small 0.38}                           &     66.83 $\pm$ {\small 0.41}     \\
\ (15) &  NST                                           & 68.35 $\pm$ {\small 0.28}                           & 67.13 $\pm$ {\small 0.35}                           &   65.60 $\pm$ {\small 0.38}       & 68.51 $\pm$ {\small 0.25}                           & 68.84 $\pm$ {\small 0.64}                           &    67.89 $\pm$ {\small 0.52}      \\
\ (16) &  RKD                                           & 65.28 $\pm$ {\small 0.81}                           & 65.63 $\pm$ {\small 0.22}                           &  63.39 $\pm$ {\small 0.52}        & 68.46 $\pm$ {\small 0.33}                           & 67.27 $\pm$ {\small 0.13}                           &     65.16 $\pm$ {\small 0.57}     \\
\midrule
\ (17) &  VBKT-GMF                                      & 69.58 $\pm$ {\small  0.49}                          & 69.96 $\pm$ {\small  0.13}                           &  \textbf{68.13 $\pm$ {\small 0.55}}        & 69.90 $\pm$ {\small  0.47}                           & 70.50 $\pm$ {\small  0.24}                           &    69.83 $\pm$ {\small  0.28}       \\
\ (18) &  VBKT-GMF-rela & \textbf{69.86 $\pm$ {\small  0.35}} & \textbf{70.23 $\pm$ {\small  0.08}} &    67.81 $\pm$ {\small 0.39}      & \textbf{70.11 $\pm$ {\small  0.33}} & \textbf{70.83 $\pm$ {\small  0.19}} &  \textbf{70.27 $\pm$ {\small  0.19}}   \\
\bottomrule
\bottomrule
\end{tabular}
\end{table*}

\subsection{Assessing VBKT-GMF Effectiveness on ASC}

First, we explore the scenario where parallel source and target device data are available, applying the proposed VBKT technique to the device adaptation task using the DCASE2020 ASC data set. Source models are initially trained on data recorded with Device A, which is referred to as the source device. The other devices are regarded as target devices.
An evaluation of the source model presented in Table \ref{tab:res-asc-base} shows that the ASC models achieve an accuracy of approximately 79\% on Device A.
However, testing those models on the eight target devices results in a significant drop in accuracy, to around 30\%, with the INCEPTION model exhibiting the lowest accuracy among them.
This experimental evidence underlines the challenge posed by the device mismatch factor.

Table \ref{tab:res-asc} presents a comprehensive comparison of various knowledge transfer methods applied to ASC across different devices.
Each entry in the table shows the average accuracy along with the standard deviation across 32 experiments, which include tests on 8 target devices conducted across 4 repeated trials for each setup.
The first three columns in Table \ref{tab:res-asc} list the results obtained using different knowledge transfer methods for the three ASC deep models without combining them with the TSL method. In contrast, the last three columns show the performance of those same methods in combination with TSL. Results with stand-alone TSL are reported in both the first and the last three columns.

With respect to the first three columns in Table \ref{tab:res-asc}`No-transfer' approach - representing models trained from scratch using the target device data only, and the `One-hot' solution - representing the source models fine-tuned on the target device data using one-hot target labels. Comparing the results of those two approaches allows us to argue that: (i) training on a small amount of target data from scratch is not effective, and (ii)  knowledge transfer is a promising solution to the device mismatch problem for the ASC task. Next, we consider the results reported from the 4th to  16th row in Table \ref{tab:res-asc} attained with advanced knowledge transfer techniques, and we can argue that: (i) most of those advanced techniques attain only a marginal improvement over the basic 'One-hot' solution, and  (ii) TSL \cite{li2014learning, hinton2015distilling}, in the fourth row,  attains a good improvement when combined with RESNET.
Finally, we consider the results delivered by the proposed variational approach which are reported in the bottom two rows in Table \ref{tab:res-asc}. We can argue that VBKT-GMF not only outperforms the `One-hot' approach by a large margin (69.58\% vs. 63.76\% for RESNET, 69.96\% vs. 64.45\% for FCNN, and 68.13\% vs. 64.38\% for INCEPTION), respectively, but it also attains a superior classification accuracy than the competing cutting-edge transfer learning techniques, which results are reported in rows 4th to 16th  in Table \ref{tab:res-asc}. Furthermore, the VBKT-GMF-rela variant, reported in the bottom row of Table \ref{tab:res-asc}, which incorporates modeling of structural relationships within the data, demonstrates a small yet consistent boost in the ASC accuracy across the three systems. Interestingly, VBKT-GMF-rela also exhibits a smaller standard deviation, indicating a more stable and reliable classification accuracy. This enhancement underscores the value of integrating structural relationships into the VBKT approach.

With respect to the last three columns in  Table \ref{tab:res-asc}. We analyze the effect of combining TSL with each of the studied transfer learning methods. Such an analysis is carried out because leveraging TSL is recommended by several studies, e.g., \cite{romero2014fitnets, zagoruyko2016paying, passalis2018learning}. Comparing, in a row-by-row manner, the results in the set of the first and the set of the last three columns, we can clearly argue that TSL always leads to a meaningful improvement in the ASC accuracy. Indeed TSL is also beneficial to our proposed  VBKT approach, and the best ASC accuracy is attained when structural relationships are considered, outperforming all the other methods considered in this work.

\subsection{Assessing VBKT-EB Effectiveness on SCR}

\begin{table}[t]
\vspace{0.3cm}
\centering
\caption{A comparison of evaluation accuracies (in \%) on target data of the SCR set. The proposed method VBKT-EB is evaluated here. Each cell of N1-N4 is averaged over 4 repeated trials, whereas the last column is the average and standard deviation of them.}
\label{tab:res-scr}
\vspace{0.1cm}
\small{
\begin{tabular}{l||cccc|c}
\toprule
\toprule
System/Domain            & N1\% & N2\% & N3\% & N4\% & Avg.\% \\
\midrule
\midrule
No-transfer        & 48.54 & 52.00 & 54.49 & 60.61 & 53.91$\pm${\footnotesize 8.98}   \\
One-hot            & 90.67 & 91.61 & 91.58 & 92.82 & 91.67$\pm${\footnotesize 0.85}  \\
\midrule
NLE           & 90.64 & 91.58 & 91.61 & 92.41 & 91.56$\pm${\footnotesize 0.75}  \\
CCKD              & 91.93 & 91.98 & 92.12 & 93.17 & 92.30$\pm${\footnotesize 0.12}  \\
NST           & 90.87 & 92.23 & 91.53 & 92.97 & 91.90$\pm${\footnotesize 0.43}  \\
PKT            & 89.21 & 90.93 & 90.06 & 91.60 & 90.45$\pm${\footnotesize 0.53}   \\
\midrule
VBKT-EB     & 92.51 & 93.19 & 92.73 & 93.80 & 93.06$\pm${\footnotesize 0.08}  \\
VBKT-EB-rela & \textbf{92.87} & \textbf{93.41} & \textbf{93.19} & \textbf{93.98} & \textbf{93.36$\pm${\footnotesize 0.07}}  \\
\bottomrule
\bottomrule
\end{tabular}
}
\end{table}

The SCR task is used to assess the VBKT-EB approach when parallel data is not available and the EB estimation method plays a crucial role. Our evaluation focuses on noise condition adaptation using the Google Speech Command data set v2, as the results are presented in Table \ref{tab:res-scr}.
It is important to note that among the knowledge transfer approaches reported in Table \ref{tab:para_setting}, only  NLE, CCKD, NST, and PKT can be evaluated on the SCR task because the other methods can be deployed only when parallel data is available.

Table \ref{tab:res-scr} reports the command recognition accuracy across four target domains, i.e., four different noisy conditions.  The last column reports the average results among the four target domains.
'No-transfer' here indicates methods that train the model with target domain noisy data from scratch. 'One-hot', which entails fine-tuning the source model using one-hot target labels and target data. As shown in the top two rows of the rightmost column, `One-hot' can improve the system performance from an average accuracy of 53.91\% for 'No-transfer' on the target domain, to an average accuracy of 91.67\% for `One-hot'. The four knowledge transfer methods reported from the 3rd to the 7th rows all can boost the SRC results, and CCKD and NST outperform `One-hot' attaining an accuracy of 92.30\% and 91.90\%, respectively. The results obtained with VBKT-EB and VBKT-EB-rela are reported in the bottom two rows in Table \ref{tab:res-scr}, respectively, and a comparison with other methods demonstrates that our solution outperforms all other competing techniques on the SCR task as well. Furthermore, leveraging structural relationships allows VBKT-EB to attain top results, as in the ASC case, and an average performance of 93.36\% is delivered.

\subsection{Effects of Hidden Embedding Depth}

\begin{figure*}[t]
     \centering
     \begin{subfigure}[b]{0.5\textwidth}
         \centering
         \includegraphics[width=\textwidth]{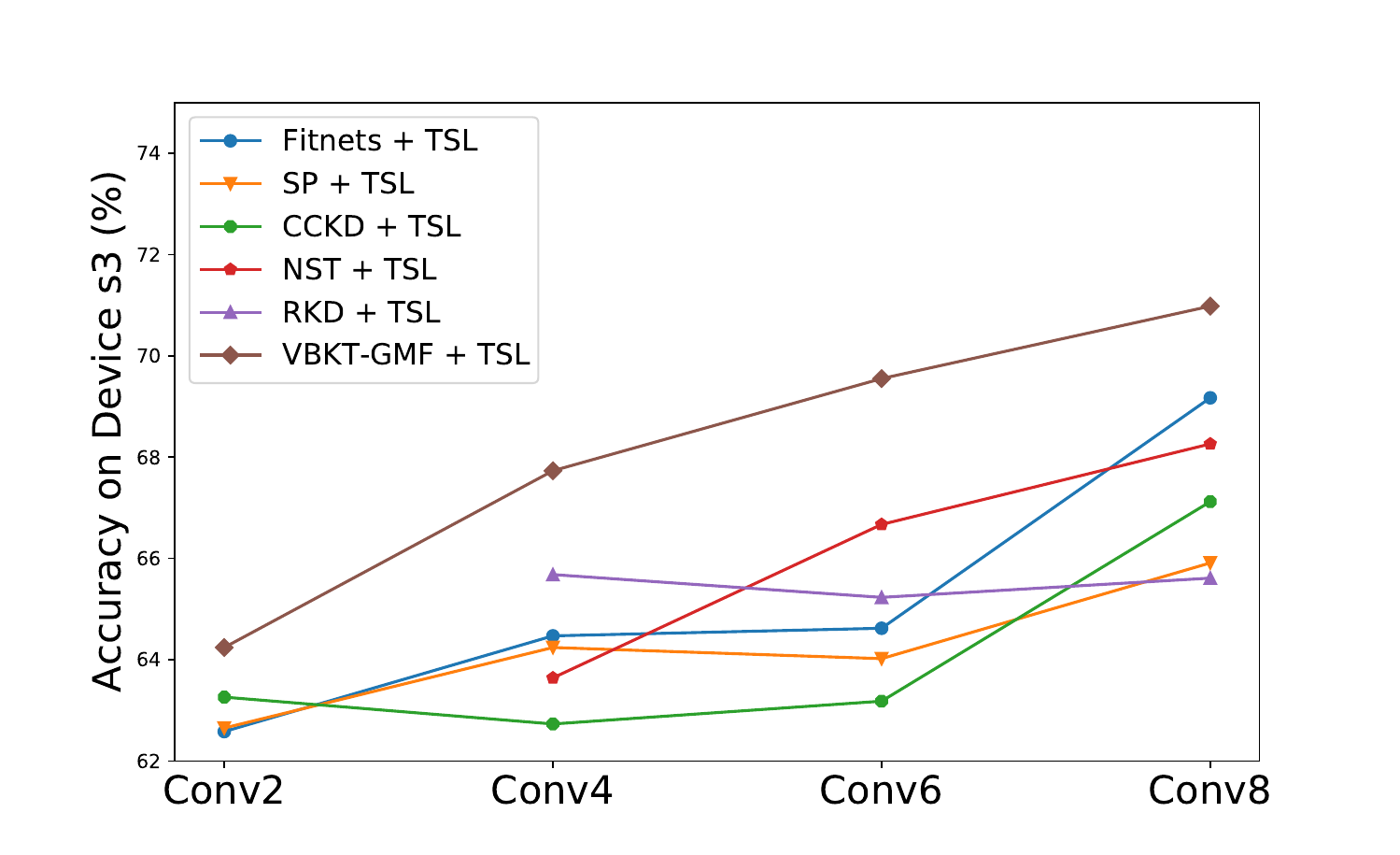}
         \caption{Device s3}
         \label{fig:res_layers_d3}
     \end{subfigure}
     \hspace{-0.4cm}
     \begin{subfigure}[b]{0.49\textwidth}
         \centering
         \includegraphics[width=\textwidth]{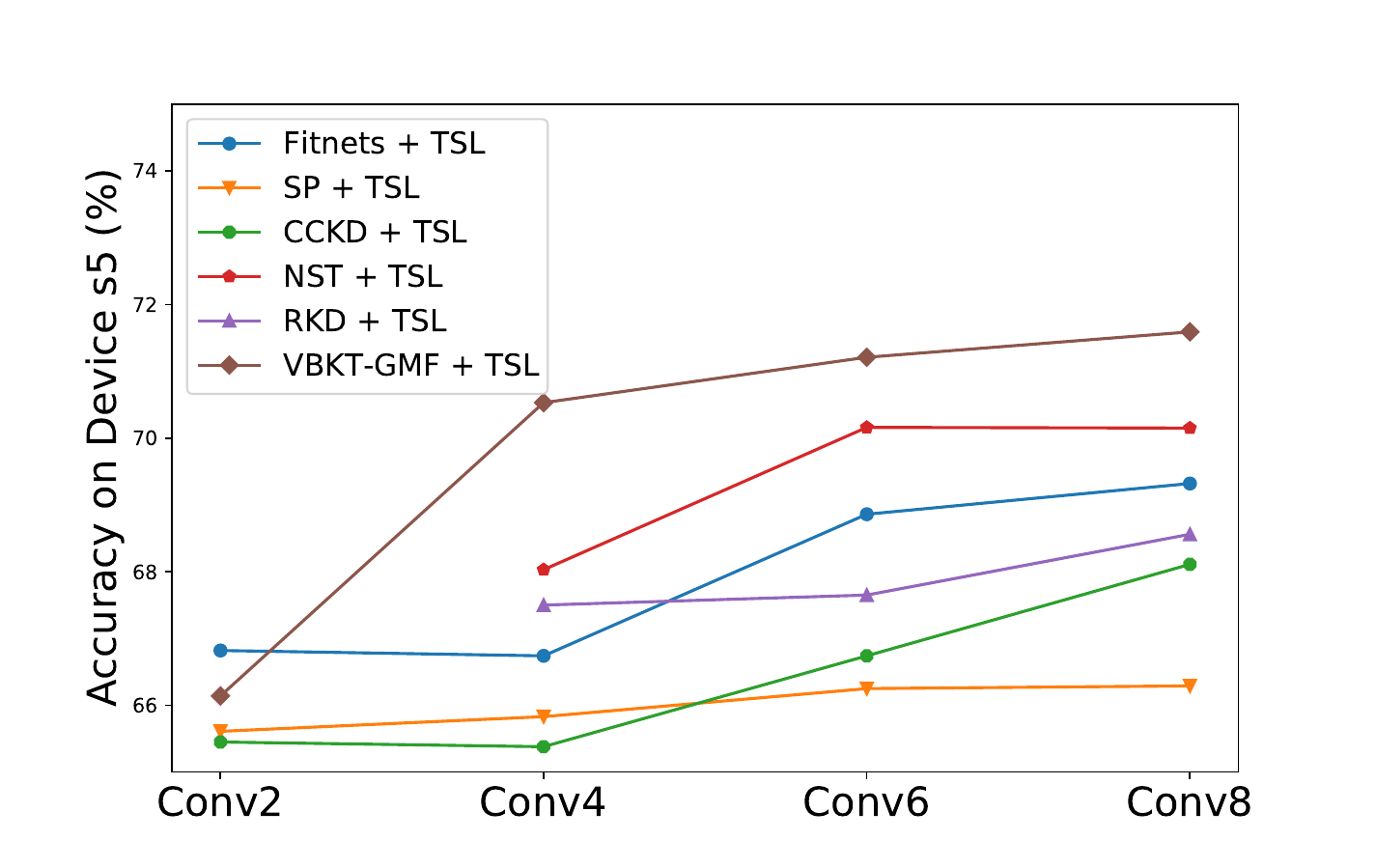}
         \caption{Device s5}
         \label{fig:res_layers_d5}
     \end{subfigure}
     \vspace{0.2cm}
        \caption{Evaluation results (classification accuracy \%) of using different hidden layers by FCNN model for ASC task. Two target devices are shown: (a) Device s3 and (b) Device s5. The basic TSL is combined with all methods.}
        \label{fig:sim}
\end{figure*}

In this study, the deep latent variables correspond to intermediate hidden embedding from specific layers within deep neural models.
In our setup, we utilize the hidden embedding from just before the final convolutional layer to model these deep latent variables. The results are depicted in Figure \ref{fig:sim}. A series of ablation experiments are conducted to explore the effect of using different layers as a hidden embedding. Experiments are performed using VBKT-GMF on two target devices of the ASC task, with the combination of TSL to obtain a better performance. The FCNN architecture is used since it has a straightforward sequential layout comprising eight stacked convolutional layers and an additional 1$\times$1 convolutional layer for outputs. We compare the hidden embedding generated by different convolutional layers from the 2nd to the 8th. Notably, models, such as COFD that utilize all hidden layers, are not considered in this comparison to isolate the effects of using specific layers.

The findings indicate that the most effective results typically emerge from using the last layer. Moreover, embedding closer to the output of the model consistently yields higher accuracies than those nearer to the input, suggesting that features from deeper layers are more effective for knowledge transfer across domains. This observation aligns with findings from other studies \cite{romero2014fitnets, huang2017like}. Finally, our proposed VBKT approach attains a very competitive ASC accuracy and consistently outperforms other methods, regardless of the hidden layer selected for analysis.

\subsection{Visualization of Intra-class Discrepancy}

\begin{figure}[t]
     \centering
     \begin{subfigure}[b]{0.15\textwidth}
         \centering
         \includegraphics[width=\textwidth]{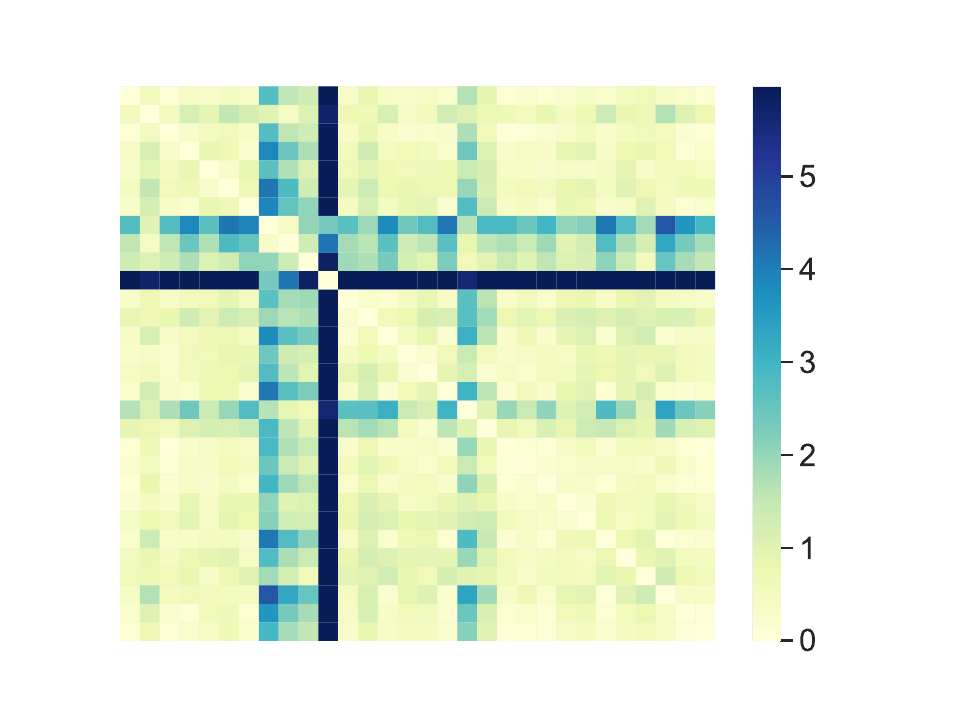}
         \caption{KD}
         \label{fig:intra_kd}
     \end{subfigure}
     \begin{subfigure}[b]{0.15\textwidth}
         \centering
         \includegraphics[width=\textwidth]{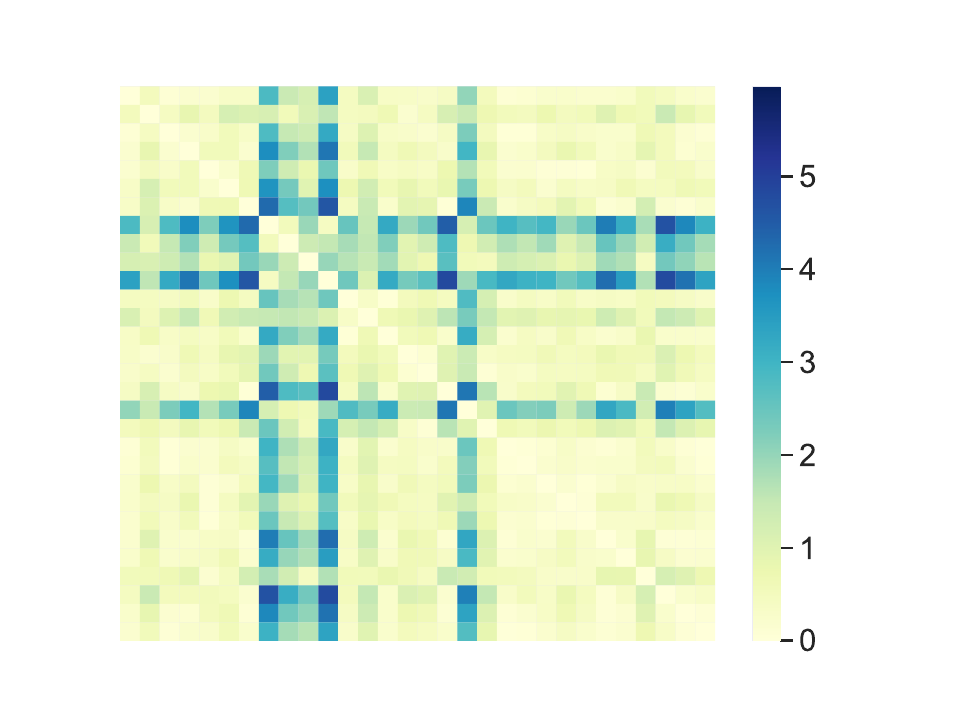}
         \caption{Fitnets}
         \label{fig:intra_fitnets}
     \end{subfigure}
     \begin{subfigure}[b]{0.15\textwidth}
         \centering
         \includegraphics[width=\textwidth]{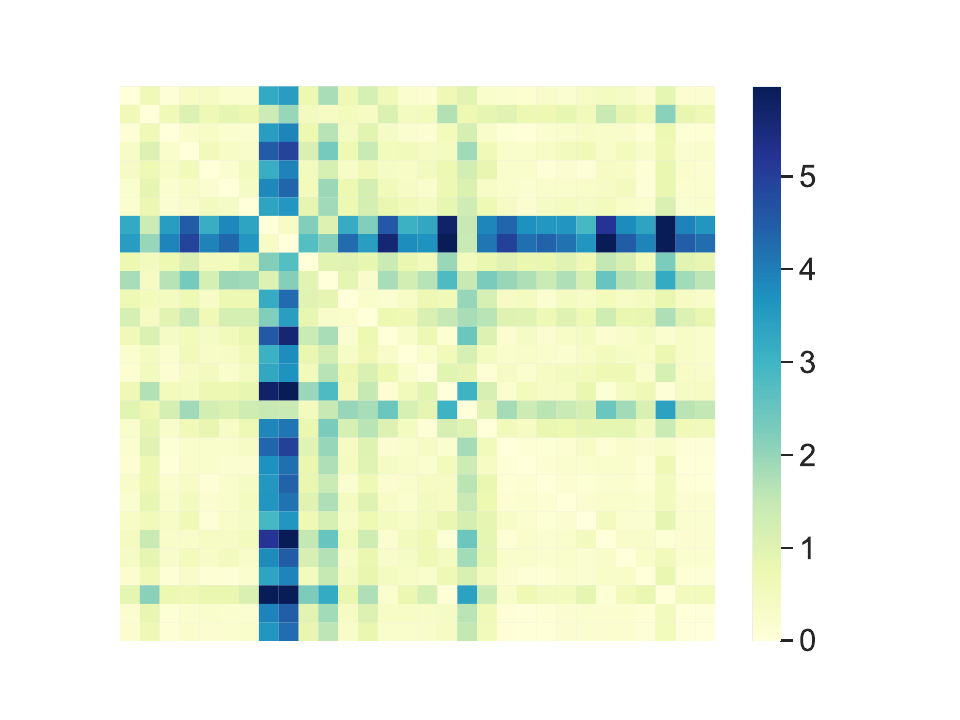}
         \caption{AT}
         \label{fig:intra_at}
     \end{subfigure}\\
     \begin{subfigure}[b]{0.15\textwidth}
         \centering
         \includegraphics[width=\textwidth]{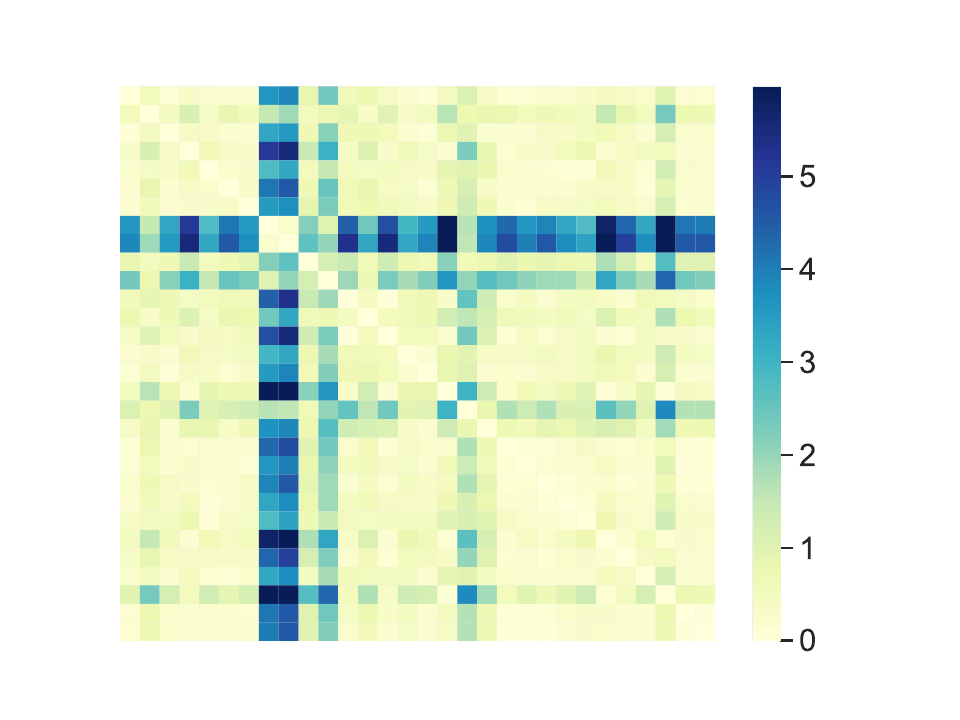}
         \caption{SP}
         \label{fig:intra_sp}
     \end{subfigure}
     \begin{subfigure}[b]{0.15\textwidth}
         \centering
         \includegraphics[width=\textwidth]{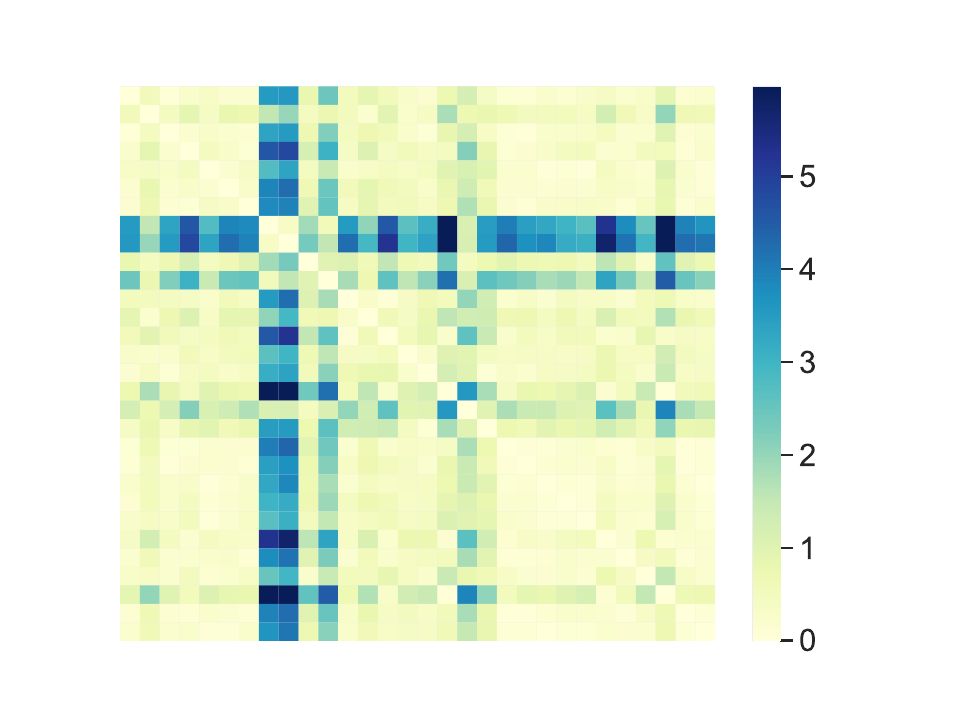}
         \caption{CCKD}
         \label{fig:intra_cckd}
     \end{subfigure}
     \begin{subfigure}[b]{0.15\textwidth}
         \centering
         \includegraphics[width=\textwidth]{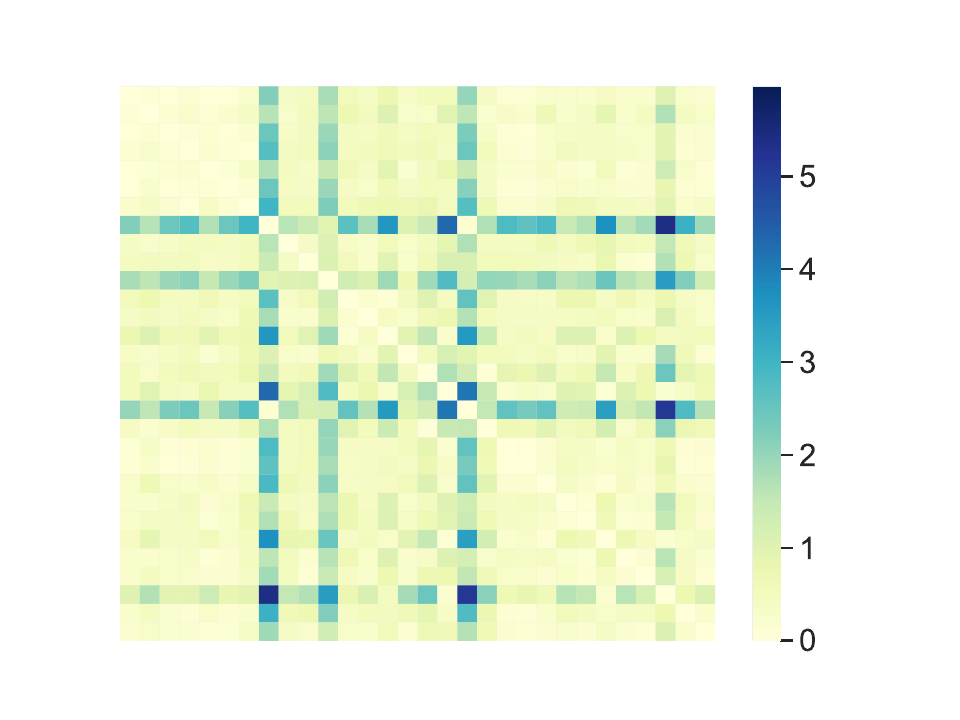}
         \caption{VBKT-GMF}
         \label{fig:intra_vbkt}
     \end{subfigure}
     \vspace{0.2cm}
    \caption{Visualized heatmaps of the intra-class discrepancy between target outputs. Each cell in the subnets illustrates the level of discrepancy between two outputs, where darker colors indicate greater intra-class discrepancies.}
    \label{fig:hetmap_intra}
\end{figure}

To better understand the effectiveness of the proposed approach, we analyze the intra-class discrepancy between the outputs of different target models prior to softmax activation.
Figure \ref{fig:hetmap_intra} displays the visualized heatmap results from (a) to (f), representing various methods evaluated on the ASC dataset. The methods include: (a) KD, (b) Fitnets, (c) AT, (d) SP, and (e) CCKD, alongside (f) our proposed VBKT-GMF approach.
For this analysis, we selected 30 random samples from the same class, where the scene class `shopping mall' is chosen as the example, and computed the $L_2$ distance between the outputs of each pair of samples. Both two axes of each subplot indicate the sample index. Each cell in the subplots of Figure \ref{fig:hetmap_intra} thus illustrates the level of discrepancy between two outputs, where darker colors indicate greater intra-class discrepancies.

From these visualization results, we can argue that the one obtained by our proposed VBKT-GMF approach in Figure \ref{fig:intra_vbkt}, exhibits consistently smaller intra-class discrepancies compared to the other methods. This observation implies that our approach enhances the discriminative capabilities of the model and significantly improves the cohesion of instances within the same class, as it implies that the model can more effectively distinguish between subtle differences inherent to different acoustic environments while maintaining robust recognition capabilities within each class.

\subsection{Feature Visualization via t-SNE}

\begin{figure}[t]
  \vspace{0.2cm}
     \centering
     \begin{subfigure}[b]{0.24\textwidth}
         \centering
         \includegraphics[width=\textwidth]{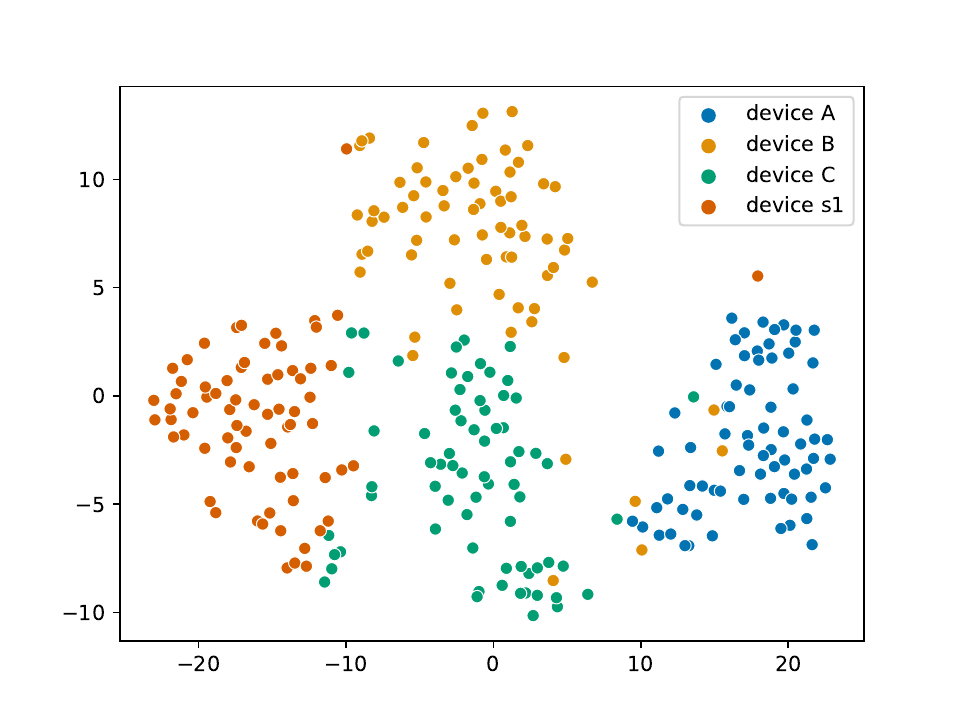}
         \caption{Source model}
         \label{fig:tsne1}
     \end{subfigure}
     \begin{subfigure}[b]{0.24\textwidth}
         \centering
         \includegraphics[width=\textwidth]{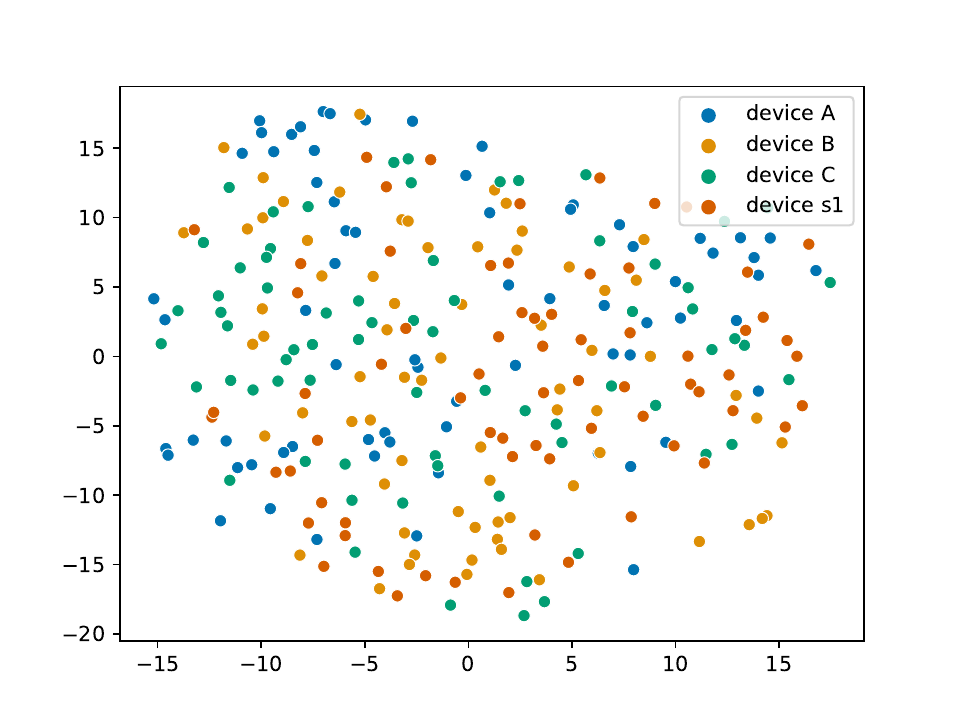}
         \caption{One-hot}
         \label{fig:tsne2}
     \end{subfigure}\\
     \begin{subfigure}[b]{0.24\textwidth}
         \centering
         \includegraphics[width=\textwidth]{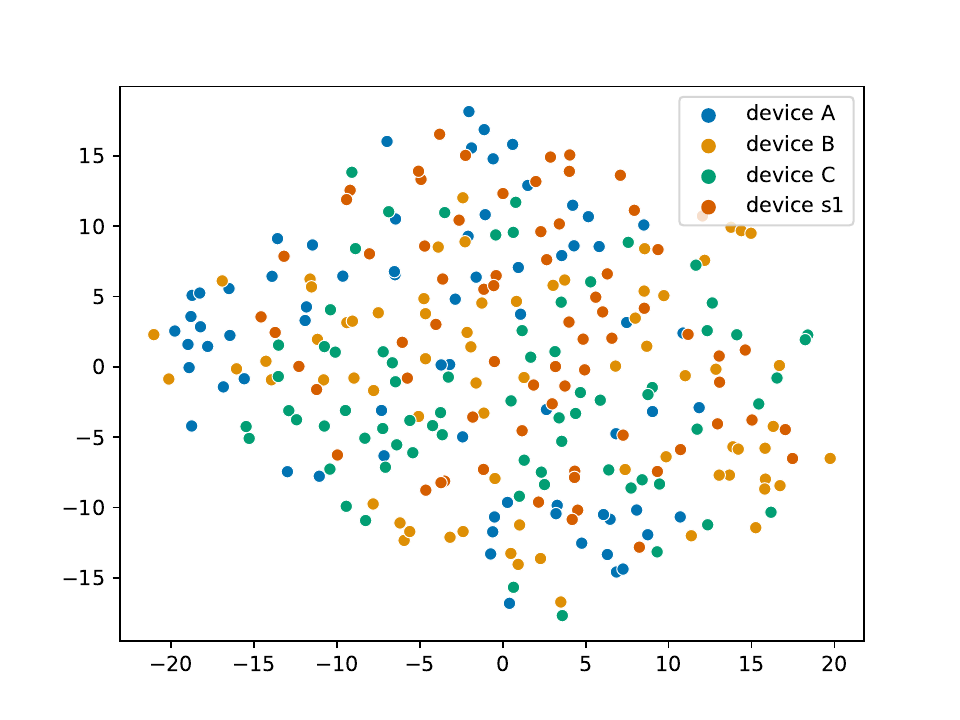}
         \caption{TSL}
         \label{fig:tsne3}
     \end{subfigure}
     \begin{subfigure}[b]{0.24\textwidth}
         \centering
         \includegraphics[width=\textwidth]{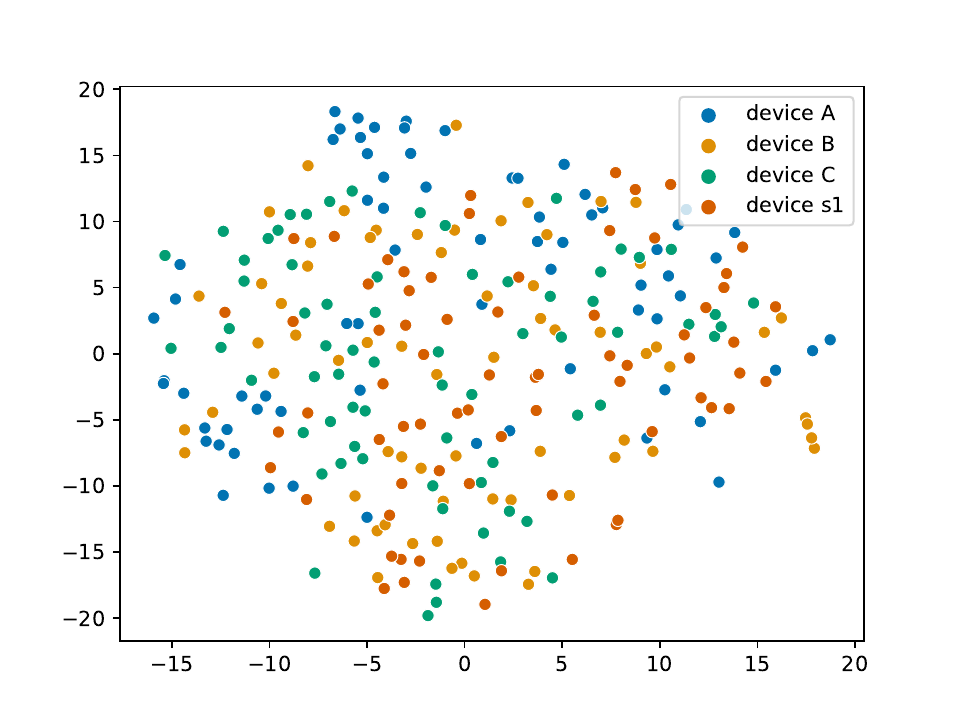}
         \caption{VBKT-GMF}
         \label{fig:tsne4}
     \end{subfigure}
     \vspace{0.2cm}
    \caption{Feature visualization via T-SNE results of four devices on ASC task. Each point in the scatter plots corresponds to a hidden feature generated from the audio sample.}
    \label{fig:tsne}
\end{figure}

We further carry out a visual analysis on ASC task. Due to the high dimensionality of the hidden embedding, we leveraged the two-dimensional t-distributed stochastic neighbor embedding (t-SNE) \cite{van2008visualizing} tool for different devices. Visualization results are shown in Figure \ref{fig:tsne} with around 300 data samples. Each point in the scatter plots corresponds to a hidden feature generated from an audio sample. In Figure \ref{fig:tsne1}, which shows the results obtained by the source model, we observe that data from different devices form distinct clusters. This is expected, as we assume that the device mismatch problem significantly affects the feature level. In contrast, in Figures \ref{fig:tsne2} to \ref{fig:tsne4}, the device-related clusters are no longer observable. We can therefore conclude that the proposed knowledge transfer technique effectively alleviates the discrepancies among devices.

\section{Conclusion}
\label{sec:con}

In this work, we introduce a novel variational Bayesian adaptive learning framework aimed at addressing cross-domain knowledge transfer challenges in acoustic applications. The core innovation of our approach lies in focusing on the estimation of deep latent variables within deep neural networks, rather than the traditional method of model parameter adaptation. This shift allows for more effective handling of acoustic mismatches arising from different scenorios such as recording devices and environmental noise. We develop two distinct estimation strategies: Gaussian mean-field variational inference (GMFVI) for scenarios with parallel source-target data and empirical Bayes (EB) for cases without such data. Additionally, we explored structural relationship modeling to enhance KL divergence approximation, preserving the structural information encoded in the latent variables.
The experimental results on device adaptation for ASC and noise adaptation for SCR task demonstrate the efficacy of our methods, with experimental results indicating consistent improvements in cross-domain adaptation performance across tasks. 
Our proposed approach can inspire future studies to explore the modeling of deep latent variables and structural priors, with different Bayesian inference methods such as MAP and MCMC, for robust acoustic adaptation.

\vspace{0.4cm}
\bibliographystyle{IEEEtran}
\bibliography{refs}

\begin{IEEEbiography}
[{\includegraphics[width=1in, height=1.25in, clip,keepaspectratio]{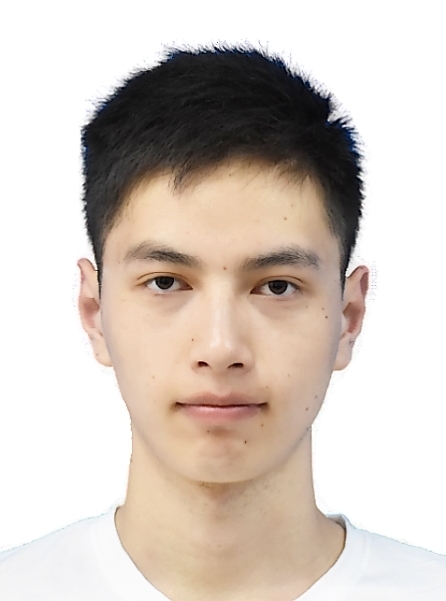}}]{Hu Hu}
received the B.S. degree from the Department of Computer Science and Technology, Shanghai Jiao Tong University, Shanghai, China, in 2018. He is currently working toward the Ph.D. degree with the School of Electrical and Computer Engineering, Georgia Institute of Technology, Atlanta, GA, USA. 
His current research focuses on Bayesian adaptation in audio and speech processing, speech recognition, and acoustic signal analysis. Besides, he was a research intern at Microsoft, Amazon, and Apple working on speech recognition and acoustic models.
\end{IEEEbiography}

\vspace{-0.5cm}
\begin{IEEEbiography}[{\includegraphics[width=1in,height=1.25in, clip,keepaspectratio]{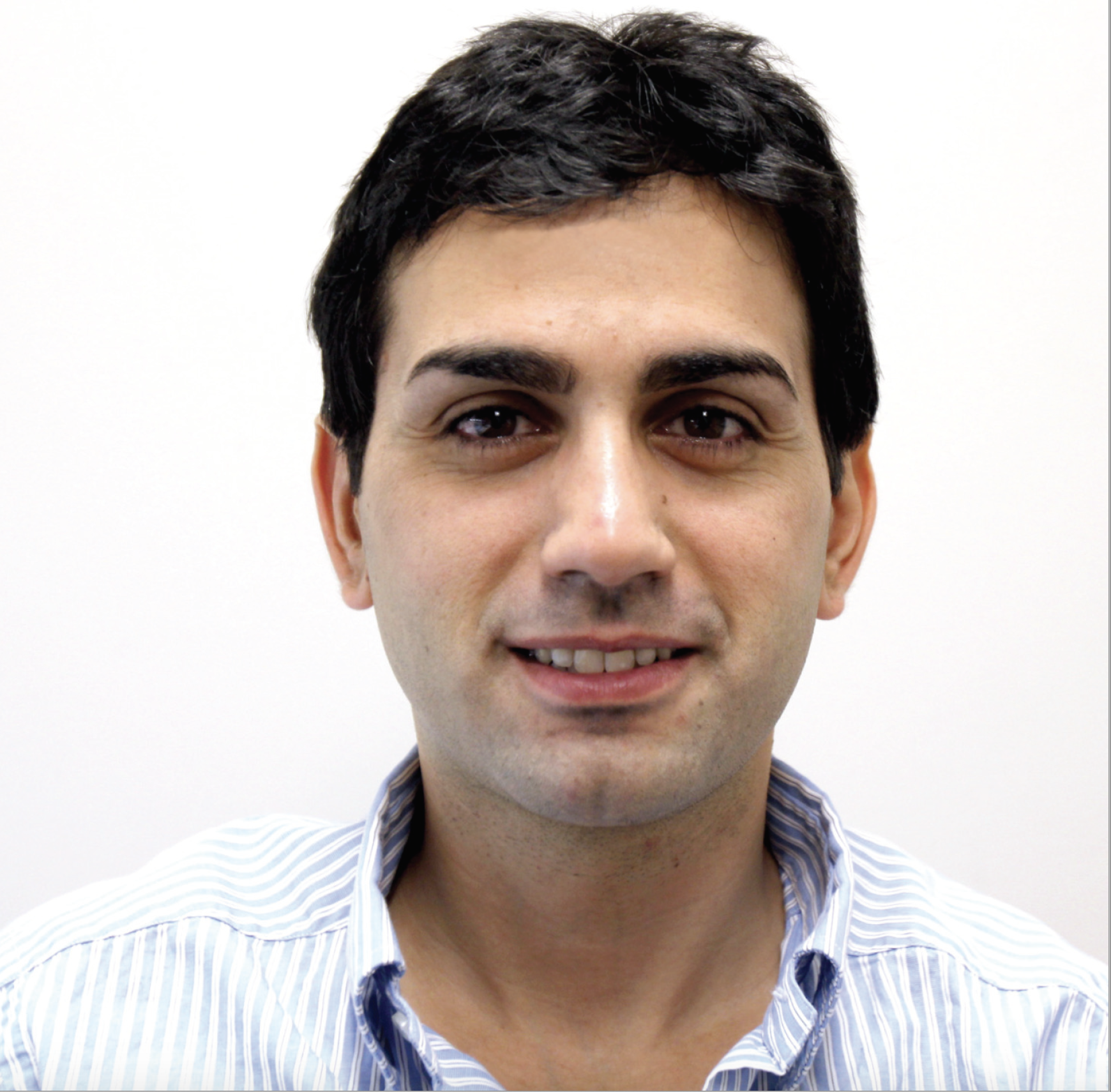}}]{Sabato Marco Siniscalchi} is a Professor at the University of Palermo, an adjunct Professor with the Norwegian University of Science and Technology, and affiliated with the Georgia Institute of Technology. He received his Laurea and Doctorate degrees in Computer  Engineering from the University of Palermo, Italy, in 2001 and 2006, respectively.  In 2006, he was a Post Doctoral Fellow at the Georgia Institute of Technology, USA. From 2007 to 2009, he joined the Norwegian University of Science and Technology (NTNU) as a Research Scientist. From 2010 to 2015, he was an Assistant Professor, first, and an Associate Professor, after, at the University of Enna. From 2017 to 2018, he joined as a Senior Speech Researcher at the Siri Speech Group, Apple Inc., Cupertino CA, USA.  In 2018, he joined the Kore University of Enna as a Full Professor. He acted as an associate editor in IEEE/ACM Transactions on Audio, Speech and Language Processing (2015-2019). He currently acts as Senior Area Editor of the IEEE Signal Processing Letters. Dr. Siniscalchi is an elected member of the IEEE Speech and Language Technical Committee (2019-2021) and (2024-2026).
\end{IEEEbiography}

\vspace{-0.5cm}
\begin{IEEEbiography}
[{\includegraphics[width=1in, height=1.25in, clip,keepaspectratio]{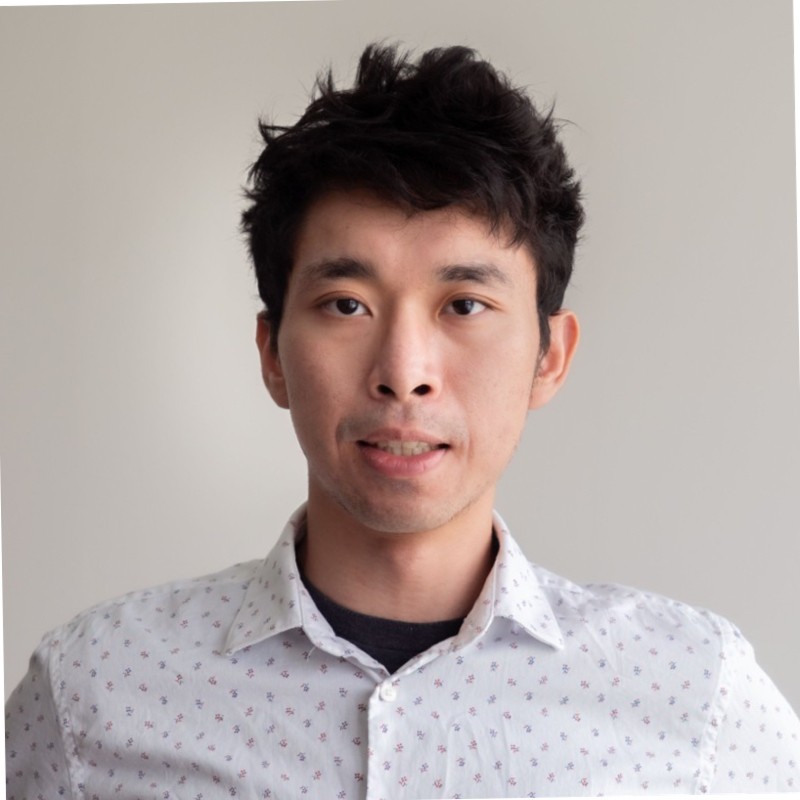}}]{Chao-Han Huck Yang} (Student Member, IEEE) received the B.S. degree from National Taiwan University, Taipei City, Taiwan, in 2016. He recently received the Ph.D. degree from the School of
Electrical and Computer Engineering, Georgia Institute of Technology, Atlanta, GA, USA, partially supported by the Wallace H. Coulter Fellowship. His research
interests include the characterizing trainable input perturbation of deep
neural networks for prompting adaptation, model robustness, privacy-preserving, and parameter-efficient learning on speech and language processing. Previously, he worked as a research intern at Google and Amazon. He recently joined NVIDIA Research as a senior research scientist.
\end{IEEEbiography}

\vspace{-0.5cm}
\begin{IEEEbiography}
[{\includegraphics[width=1in, height=1.25in, clip,keepaspectratio]{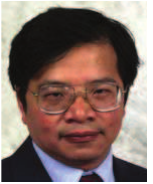}}]{Chin-Hui Lee}
(Fellow, IEEE) is a Professor with the School of Electrical and Computer Engineering, Georgia Institute of Technology. Before joining academia in 2001, he had 20 years of industrial experience, ending at Bell Laboratories, Murray Hill, NJ, USA, as a Distinguished Member of Technical Staff, and the Director of the Dialogue Systems Research Department. He has authored or coauthored more than 600 papers and patents. They have been more than 55,000 times with an $h$-index of 80 on Google Scholar. He has received numerous awards, including the Bell Labs President's Gold Award in 1998. He also won the 2006 Technical Achievement Award from IEEE Signal Processing Society for Exceptional Contributions to the Field of Automatic Speech Recognition. In 2012, he was invited by ICASSP to give a plenary talk on the future of speech recognition. In the same year, he was awarded the ISCA Medal for Scientific Achievement for "Pioneering and Seminal Contributions to the Principles and Practices of Automatic Speech and Speaker Recognition".
\end{IEEEbiography}

\end{document}